\documentclass[aps,english,twoside,twocolumn,prb]{revtex4}

\usepackage{amsmath}
\usepackage{graphicx}
\usepackage{amssymb}
\usepackage{babel}
\usepackage{color}

\makeatother
\makeatletter

\newcommand{\noun}[1]{\textsc{#1}}

\begin{document}

\title{Adiabatic dynamics of a quantum critical system coupled to an environment:\\
Scaling and kinetic equation approaches}

\author{Dario Patan\`{e}}
\affiliation{MATIS CNR-INFM $\&$ Dipartimento di Metodologie Fisiche e Chimiche (DMFCI),
Universit\`a di Catania, viale A. Doria 6, 95125 Catania, Italy}
\affiliation{Departamento de Física de Materiales, Universitad Complutense, $28040$
Madrid, Spain}

\author{Alessandro Silva}
\affiliation{The Abdus Salam International Centre for Theoretical Physics, Strada
Costiera $11$, $34100$ Trieste, Italy }

\author{Luigi Amico}
\affiliation{MATIS CNR-INFM $\&$ Dipartimento di Metodologie Fisiche e Chimiche (DMFCI),
Universit\`a di Catania, viale A. Doria 6, 95125 Catania, Italy}
\affiliation{Departemento de Física de Materiales, Universitad Complutense, $28040$
Madrid, Spain}

\author{Rosario Fazio}
\affiliation{NEST-CNR-INFM $\&$ Scuola Normale Superiore, Piazza dei Cavalieri 7,
I-56126 Pisa, Italy}

\author{Giuseppe E. Santoro}
\affiliation{International School for Advanced Studies (SISSA), Via Beirut $2-4$,
$34014$ Trieste, Italy }
\affiliation{CNR-INFM Democritos National Simulation Center, Via Beirut $2-4$,
$34014$ Trieste, Italy }
\affiliation{The Abdus Salam International Centre for Theoretical Physics, Strada
Costiera $11$, $34100$ Trieste, Italy }

\begin{abstract}
We study the dynamics of open quantum many-body systems driven across
a critical point by quenching an Hamiltonian parameter at a certain
velocity. General scaling laws are derived for the density of excitations
and energy produced during the quench as a function of quench velocity
and bath temperature. The scaling laws and their regimes of validity
are verified for the XY spin chain locally coupled to bosonic baths.
A detailed derivation and analysis of the kinetic equation of the
problem is presented.
\end{abstract}

\maketitle

\section{Introduction}

A series of beautiful experiments  on the dynamics  of cold atomic
gases~\cite{Greiner,Kinoshita,Sadler} spurred renewed interest in
the study  of non-equilibrium quantum many-body systems. On the
theoretical side these experiments triggered an intense
investigation mostly devoted to the simplest paradigm of
nonequilibrium quantum dynamics: the controlled variation in time of
one of the system parameters (quantum quenches). In the case of
sudden quenches, where the driving parameter is changed on a time
scale much shorter than typical time scales of the system, a number
of important issues have been addressed. We mention, for example,
the study of the signatures of universality in the quench dynamics
of quantum critical systems~\cite{Sengupta}, the presence of
thermalization in integrable vs. nonintegrable systems~\cite{Rigol},
as well as the description of generic nonequilibrium quenches using
thermodynamic variables ~\cite{Polkovnikov08-2} and their
statistics~\cite{Silva08}.

In this paper we will focus on the opposite case in which the
control parameter is varied {\em adiabatically}, a case which
becomes particularly interesting when a critical point is crossed
during the adiabatic evolution. Because of the vanishing of the
energy gap at criticality, the system is unable to follow
adiabatically the driving remaining in its equilibrium/ground state
when passing through the quantum critical point the system will not
be able  no matter how slow is the quench. The study of these
deviation from the adiabatic dynamics is a problem which is very
important in a number of different branches of physics ranging from
the defect formation in the early universe~\cite{KZ,KibbleReview} to
adiabatic quantum computation~ \cite{farhi01} or quantum
annealing\cite{santoro02,santoro06}. Depending on the context, the
loss of adiabaticity has been characterized by the excess energy at
the end of the quench, by the density of defects (if the final state
was a fully ordered system), or by the fidelity of the time evolved
state with the ground state of the Hamiltonian at the end of the
quench.

The scaling of the density of excitations generated during the
dynamics as a function of the velocity of the quench was first
predicted in Ref.~\onlinecite{zurek05,polkovnikov05} for a quantum
critical system. The mechanism behind the generation of
excitations/defects is  similar to so-called Kibble-Zurek (KZ)
mechanism~\cite{KZ} first proposed for classical phase transitions.
Following these initial works a number of specific models were
scrutinized\cite{Dziarmaga05,Damski05,
Schutzhold06,Cherng06,Damski07,Cucchietti07,Cincio07,Caneva07,Sengupta08,Polkovnikov08-1,Caneva08,Deng08,
Sen08,Pellegrini08,Divakaran08}, thereby confirming the general
picture.

All the works mentioned previously assumed unitary Hamiltonian
dynamics. We know, however, that understanding the effect of the
external environment on the adiabatic dynamics is of paramount
importance for several reasons. In the case of adiabatic quantum
computation, decoherence is a fundamental limiting factor to the
ability of implementing quantum algorithms.  Furthermore, an
experimental verification of the KZ scaling in a quantum phase
transition can only  occur through the detection of this effect at
low temperatures, i.e. when the quantum critical system is in
contact with a thermal bath. Despite its importance the adiabatic
dynamics of open critical systems is a much less studied problem.
The effect of classical and quantum noise  acting uniformly on a
quantum Ising chain was considered in Ref.\onlinecite{Fubini07} and
Ref.\onlinecite{Mostame07} respectively.
Numerical simulations for a model of  local noise on a disordered 
Ising model were performed in Ref.\onlinecite{Amin08}. 
Moreover the effect a static spin bath locally coupled to an ordered
Ising model is studied  in Ref.\onlinecite{Cincio08}. 
In a recent Letter~\cite{PatanePRL}, we
have addressed the universality of the production of defects in the
adiabatic  dynamics in the presence of an environment by
generalizing the scaling theory to open critical system and by
formulating a quantum kinetic equation approach for the adiabatic
dynamics across the quantum critical region. We found that, at weak
coupling and for not too slow quenches the density of excitations is
universal also in the presence of an external bath. In this paper we
extend the results presented in Ref.~\onlinecite{PatanePRL} and
provide a detailed derivation of the kinetic equations and of the
scaling approach.

The paper is organized as follows.  We first derive qualitatively
the scaling laws obeyed by both the density of defects and  of
energy generated in a quench for a generic open quantum critical
system (Sec.\ref{sec:Scaling-analysis}). We then address  a specific
one-dimensional model possessing a quantum critical point: the XY
spin chain in transverse magnetic field.  To model a thermal
reservoir we couple the system to a set of bosonic degrees of
freedom, as in the spin-boson model. Baths are chosen with power-law
spectral density and are locally coupled to strings of neighboring
spins. The model, a generalization of the one studied in
Ref.\onlinecite{PatanePRL}, is discussed in  Sec.\ref{Model}. For
this model, we derive a kinetic equation within the Keldysh
technique (Sec. \ref{sec:Kinetic-equation} and Appendix A1-A2) which
allows us to compute the density of defects. In  Sec.
\ref{sec:Relaxation-time} (and Appendix B) we discuss the spectrum
of relaxation times needed for a comparison with the scaling
approach. The density of defects and of excitation generated in a
quench, and a comparison with the scaling laws is presented in Sec.
\ref{sec:Adiabatic-quenches}. Finally in Sec.\ref{conclusions} we
summarize our conclusions.

\section{Scaling analysis} \label{sec:Scaling-analysis}

\begin{figure}[h]
\includegraphics[scale=0.42]{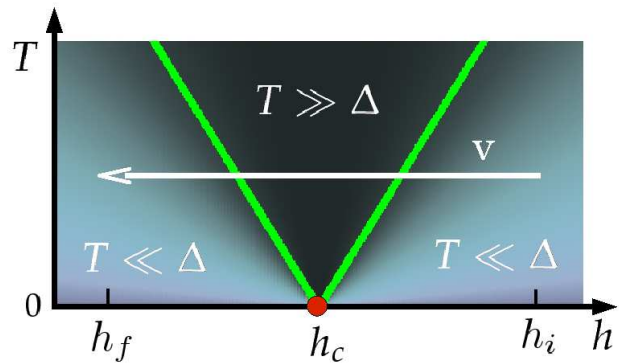}
\caption{\label{fig:fig1} A sketch of the finite temperature crossover phase-diagram
close to the quantum critical point. Crossover lines $T\sim|h-h_{c}|^{\nu z}$
separating the semiclassical regions from the quantum critical region are shown.
The latter is traversed by the system during the quench in a time $t_{QC}$.}
\end{figure}

In this section we discuss the scaling laws obeyed by the density of
excitations \cite{PatanePRL}  and by the energy density following a
linear quench of  a control parameter $h$ from an initial value
$h_i$ to a final one $h_f$,  through a second-order quantum critical
point at $h_c$.  The system, during the whole dynamics, is kept in
contact with a bath at temperature $T$.  In the $h-T$ plane the
adiabatic quench is described by the horizontal line shown in
Fig.\ref{fig:fig1}. For adiabatic quenches occurring at zero
temperature, the system stops following the external drive
adiabatically and can be considered as frozen around the quantum
critical point. This happens roughly when the time it takes to reach
and cross the quantum critical point becomes comparable to the
internal time scale (the inverse gap $\Delta(h) \simeq |h-h_c|^{-\nu
z}$). The determination of this crossover point is the fundamental
ingredient which leads to the scaling in the case of unitary
evolution~\cite{zurek05,polkovnikov05}. In the case of a finite
temperature quench there is a new important timescale which enters
the problem, the time at which the system enters (and eventually
leaves) the quantum critical region (see Fig.\ref{fig:fig1}).
Initially the system is in equilibrium with the bath at a low
temperature ($T\ll\Delta(h_{i})$) and the behavior is, to a large
extent,  as in the zero-temperature case. In the quantum critical
region~\cite{SachdevBOOK}, characterized by the crossover
temperature  $T\sim|h-h_{c}|^{\nu z}$, the gap is much smaller than
the temperature itself. One can therefore expect that during the
interval the system spends in the quantum critical region a number
of excitations will be produced by the presence of the environment.
Interestingly also this  contribution to the defect production obeys
a scaling law~\cite{footnote1}.

We now proceed with the derivation of the scaling laws. In the rest of the paper we will consider the density
$\mathcal{E}$ and the energy density ${E}$ of excitations, defined respectively:
\begin{eqnarray}
\mathcal{E} &=& \int \frac{d^{d}k}{\left(2\pi\right)^{d}} \mathcal{P}_{k} \;,
\label{eq: excitations density}\\
{E}  &=& \int \frac{d^{d}k}{\left(2\pi\right)^{d}} E_{k} \mathcal{P}_{k} \;,
\label{eq:eenergy density}
\end{eqnarray}
where $\mathcal{P}_{k}$ is the population of the excitation with quantum number $k$,
$E_{k}$ the energy spectrum at $h_f$, and $d$ is the dimensionality of the quantum system.

The first assumption we make consists in separating the density of excitations/energy at the end of the quench
in the sum of two (coherent and incoherent) contributions:
\begin{eqnarray}
\mathcal{E} &\simeq& \mathcal{E}_{coh}+\mathcal{E}_{inc} \;, \label{eq:ebath+eclean}\\
{E}  &\simeq& {E}_{coh}+{E}_{inc}  \;. \label{eq:etabath+etaclean}
\end{eqnarray}
In the previous equation,  ${E}_{coh}$ ($\mathcal{E}_{coh}$) is the
density of energy (excitations) of the system produced coherently in
the absence of the bath; the incoherent contribution ${E}_{inc}$
($\mathcal{E}_{inc}$) arises instead from the bath/system
interaction. The separation of a coherent and an incoherent
contribution, eqs.~(\ref{eq:ebath+eclean}) and
(\ref{eq:etabath+etaclean}), requires weak  coupling $\alpha$
between the system and the bath. Evidence for the validity of this
assumption will be shown below (see Eqs.(\ref{eq:Ebath_generic_QPT})
and (\ref{eq:ETAbath_genericQPT})).

In the absence of an environment, the density of excitations was
shown to obey the KZ scaling \cite{zurek05,polkovnikov05}
\begin{equation}
\mathcal{E}_{coh} = \mathcal{E}_{KZ} \propto v^{d\nu/(z\nu+1)} \;.
\label{eq: E KZ}
\end{equation}
In order to obtain a similar relation for the energy density in Eq.~(\ref{eq:energy density})
additional information on $E_k$ at $h_{f}$ is needed.
Thus, the scaling of this quantity depends on the details of the
system at the end of the quench. A simple scaling law can be
obtained only in specific situations, e.g. for quenches halted at
the critical point $h_{f}=h_{c}$, where $E_{k}\propto k^{z}$.
By using techniques similar to those employed in Ref.~\onlinecite{polkovnikov05}
one obtains
\begin{equation}
{E}_{coh} \propto v^{\nu(d+z)/(z\nu+1)} \;.
\label{eq: eta KZ}
\end{equation}

Let us now derive a scaling law for the incoherent contributions $\mathcal{E}_{inc}$ and ${E}_{inc}$.  To this end it is
convenient to divide the quench in three steps (see Fig.~\ref{fig:fig1}):  initially the system is in the so-called low-temperature
region at $T\ll\Delta$.  Here the relatively high energy gap suppresses thermal excitation and the system remains in the
ground state.  Close to the critical point the system passes through the quantum critical region: thermal excitations are
unavoidably created because of the relatively high temperature $T\gg\Delta$.  As we shall see below, the density of excitations
generated in this region are universal functions on the velocity of the quench and on the temperature as long as only the low-energy
details of the system spectrum matter.  On the contrary, the bath-induced relaxation occurring once the system leaves
the quantum critical region, entering the other semiclassical region ($T\ll\Delta$), depends on the details of the energy spectrum;
hence the relaxation towards an asymptotic thermal state at temperature $T$ is not expected to be universal if the final $h_{f}$ is far
off the critical point $h_c$.  In our analysis below we will neglect the effects of this non-universal relaxation.
We are therefore assuming that the time elapsed between the moment when the critical region is left and
when the quench is stopped (and the measurement of the density of excitations/energy is made) is short  as compared to the typical
relaxation times in the semiclassical region.  In Section \ref{sec:Adiabatic-quenches} we will further comment on this point for the specific
case of the quantum XY chain, showing that the scenario just depicted holds for a wide range of $h_{f}$ and $v$.
The dynamics of the probability to excite the model $k$  $\mathcal{P}_{k}$  can be described, inside the quantum critical region,  in
terms of a phenomenological rate equation:
\begin{equation}
\frac{d}{dt} \mathcal{P}_{k}=- \frac{1}{\tau} \left[ \mathcal{P}_{k}-\mathcal{P}_{k}^{th}\left(h_{c}\right) \right] \;,
\label{eq:rate eq P}
\end{equation}
where $\mathcal{P}_{k}^{th}(h_{c})$ is the critical thermal equilibrium distribution
and $\tau^{-1}$ is the relaxation rate, $\tau^{-1}\propto\alpha T^{\theta}$.
As shown in Section \ref{sec:Relaxation-time} and appendix C, $\theta$
can be related to characteristics of the bath and to the critical
indices of the phase transition (see Eq. (\ref{eq:tau general appendix})).
From the relation $T\sim\Delta\sim |h-h_{c}|^{\nu z}$ we deduce the time spent inside the quantum critical
region is
$$t_{QC}=2T^{1/\nu z}v^{-1} \;\;. $$
A direct integration of Eq.~(\ref{eq:rate eq P}) gives for the thermal excitation created in the quantum critical region
$\mathcal{P}_{k} \sim (1-e^{t_{QC}/\tau}) \mathcal{P}_{k}^{th} \left(h_{c}\right)$.
Finally, integrating the latter over all k-modes we get:
\begin{eqnarray}
\mathcal{E}_{inc} &\propto& \left(1-e^{t_{QC}/\tau}\right)
\int \! dE \, E^{d/z-1}\, \mathcal{P}_{k}^{th}(h_{c}) \;,
\label{eq: Einc int}
\end{eqnarray}
where we used the scaling of the excitation  energy $E\propto k^{z}$.
For the density of energy, a similar relation holds in the case of
quenches halted at the critical point
\begin{eqnarray}
{E}_{inc} &\propto& \left(1-e^{t_{QC}/(2\tau)}\right) \int \! dE \, E^{d/z} \, \mathcal{P}_{k}^{th}(h_{c}) \;,
\label{eq:ETAinc int}
\end{eqnarray}
where $t_{QC}/2$ is due to the fact that in this case only half of the
quantum critical region is crossed. Finally, since the thermal distribution
is a function of $E/T$, a simple change of variable gives the required results
\begin{eqnarray}
\mathcal{E}_{inc} &\propto& \alpha \, v^{-1}\, T^{\theta+\frac{d\nu+1}{\nu z}} \;,
\label{eq:Ebath_generic_QPT}\\
{E}_{inc} &\propto& \alpha \, v^{-1}\, T^{\theta+\frac{(d+z)\nu+1}{\nu z}} \;,
\label{eq:ETAbath_genericQPT}
\end{eqnarray}
that are valid in the limit $T^{1/\nu z}\ll v\tau$.
Eqs.~(\ref{eq:Ebath_generic_QPT}) and (\ref{eq:ETAbath_genericQPT})
together with Eqs.~(\ref{eq: E KZ}) and (\ref{eq: eta KZ}) give, through
the assumptions Eqs.~(\ref{eq:ebath+eclean}) and (\ref{eq:etabath+etaclean}),
the general scaling-law for the quench dynamics of open systems.
The different scaling of the two contributions with respect to the velocity $v$
implies that for slow quenches $v<v_{cross}$ the incoherent mechanism of excitation dominates
over the coherent one, and viceversa for $v>v_{cross}$.
The crossover velocity can be deduced by equating $\mathcal{E}_{inc}\simeq\mathcal{E}_{coh}$
and ${E}_{inc}\simeq {E}_{coh}$ yielding:
\begin{eqnarray}
v_{cross}^{\mathcal{E}} &\propto& \alpha^{\frac{\nu z+1}{\nu(z+d)+1}}\,
T^{\left(1+\frac{(\theta-1)\nu z}{\nu(z+d)+1}\right)\left(1+\frac{1}{\nu z}\right)} \;,
\label{eq:E vbath_generic_QPT}\\
v_{cross}^{E} &\propto& \alpha^{\frac{\nu z+1}{\nu(2z+d)+1}}\,
T^{\left(1+\frac{(\theta-1)\nu z}{\nu(2z+d)+1}\right)\left(1+\frac{1}{\nu z}\right)} \;.
\label{eq:ETA vbath_generic_QPT}
\end{eqnarray}
%

\section{Quantum XY model with thermal reservoir} \label{Model}

The scaling laws derived above will be tested against a specific model: an XY-chain coupled to a
set of bosonic baths.
The Hamiltonian of the XY chain is defined as
\begin{equation}
H_{S}=-\frac{1}{2} \! \sum_{j}^{N} \!
\left(\frac{1+\gamma}{2}\sigma_{j}^{x}\sigma_{j+1}^{x}+\frac{1-\gamma}{2}\sigma_{j}^{y}\sigma_{j+1}^{y}
\!+\!  h\sigma_{j}^{z}\right) \;.
\label{eq: H_S}
\end{equation}
Here $N$ is the number of sites ($\sigma^{x,\ y,\ z}$ are Pauli matrices). Each spin is coupled to its neighbors d
by anisotropic Ising-like interaction and subject to a transverse magnetic field $h$ (the couplings are expressed in terms of
the exchange energy).
In the thermodynamic limit $N\rightarrow\infty$, a quantum phase transition at $h_{c}=1$ separates
a paramagnetic phase for $h>1$ from a ferromagnetic phase ($h<1$) where the $Z_{2}$ symmetry is spontaneously
broken and a magnetic order along $\vec{x}$ appears, $\left\langle \sigma^{x}\right\rangle \ne 0$.

The spin Hamiltonian Eq.i~(\ref{eq: H_S}) can be diagonalized by using the Jordan-Wigner transformation~\cite{pfeuty}
to map the spins into spinless fermions $c_{j}$ and thus obtain in momentum space
(after a projection in a definite parity subspace)
\begin{eqnarray}
H_{S} &=& \sum_{k>0}\Psi_{k}^{\dagger}\hat{\mathcal{H}}_{k}\Psi_{k}\nonumber \\
\hat{\mathcal{H}}_{k} &=& -(\cos k+h)\, \hat{\tau}_{z} + \gamma\sin k \, \hat{\tau}_{y} \;,
\label{eq: H matrix Nambu}
\end{eqnarray}
where $\Psi_{k}^{\dagger}=\left(\begin{array}{cc} c_{k}^{\dagger} & c_{-k}\end{array}\right)$
are Nambu spinors and $\hat{\tau}$ are Pauli matrices in Nambu space.
Finally, a Bogoliubov rotation diagonalizes the Hamiltonian:
$H_{S}=\sum_{k>0}\Lambda_{k}(\eta_{k}^{\dagger}\eta_{k}-\eta_{-k}\eta_{-k}^{\dagger})$,
where
\begin{equation}
\Lambda_{k}=\sqrt{(\cos k+h)^{2}+(\gamma\sin k)^{2}} \;,
\label{eq:energy spectrum}
\end{equation}
is the quasi-particle dispersion. At $h=h_{c}$ the spectrum becomes gapless with a
linear dispersion relation $\Lambda_{k}\propto\pi-k$; 
accordingly, the critical indexes of the model are $\nu=z=1$.

The spins are also  locally coupled to a set of $N/l$ bosonic baths
\begin{equation}
H_{int}=-\frac{1}{2} \sum_{j=0}^{N/l-1} \left(\sum_{r=0}^{l-1}\sigma_{jl+r}^{z} \right) X_{j}
\label{eq:H_int}
\end{equation}
where $X_{j}=\sum_{\beta} \lambda_{\beta} (b_{\beta,j}^{\dagger}+b_{-\beta,j})$
and $b_{\beta,j}^{\dagger}$ ($b_{\beta,j}$) are the creation (annihilation)
operators of the j-th bosonic bath. As a result of the coupling in Eq.~(\ref{eq:H_int})
each baths correlated in a string of $l$ adjacent spins.
The model presented above generalizes the one considered in Ref.~\onlinecite{PatanePRL},
where the spins were individually coupled ($l=1$) to Ohmic baths ($s=1$).

The total Hamiltonian reads:
\begin{equation}
H=H_{S}+H_{int}+H_{B}
\label{eq:H_TOT}
\end{equation}
where $H_{B}=\!\sum_{j,\beta}\!\omega_{\beta} b_{\beta j}^{\dagger} b_{\beta j}$.
The spectral density of the baths $J(\omega)=\sum_{\beta} \lambda_{\beta}^{2} \delta(\omega-\omega_{\beta})$ is
\begin{equation}
J(\omega) = 2\alpha\omega^{s} e^{-\omega/\omega_{c}} \theta(\omega)
\label{eq:Bath Spectral Function}
\end{equation}
where $\alpha$ is the system/bath coupling, $\omega_{c}$ is a high energy cutoff and $\theta(s)$ is the
step function \cite{WeissBOOK}.

\begin{figure}[h]
\includegraphics[scale=0.4]{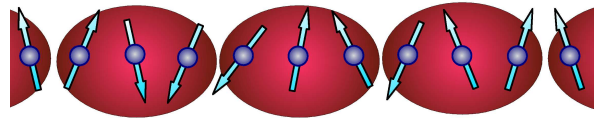}
\caption{\label{fig:fig2} A cartoon of the spins-baths coupling (\ref{eq:H_int}) for the $l=3$,
where each bath is coupled to three spins.}
\end{figure}

In the  situation we are interested, the system is initialized in its ground state at large $h$. The coupling of the
spin to the bath through $\sigma^{z}$ preserves parity symmetry (see Eq.~(\ref{eq: H_int_NAMBU})).
Therefore, once the system has initially a specific parity, it will remain in the
corresponding sector for the entire evolution. Throughout this paper we consider $N$ even:
in this case the ground state has always an even number of fermions $c_{k}$ and we are thus allowed to select
the even parity sector and neglect the odd one.

In momentum space
$b_{\beta,q}=\frac{1}{\sqrt{N/l}}\sum_{j=0}^{N/l-1}\exp(-iqj) \, b_{\beta,j}$
with $q=\frac{2m\pi}{N/l}$ and after the Jordan-Wigner transformation we get
\begin{equation}
H_{int} =-\frac{1}{\sqrt{N}} \sum_{k} \sum_{q} F(q) \Psi_{k}^{\dagger}\hat{\tau}^{z} \Psi_{k+\frac{q}{l}} X_{q}
\label{eq: H_int_NAMBU}
\end{equation}
where $F(q)=1/\sqrt{l} \sum_{r=0}^{l-1} \exp(-i r q/l)$.
For $l=1$ each spin interact with a different bath and according to Eq.~(\ref{eq: H_int_NAMBU})
all $k$ modes are coupled (i.e., transitions $k \leftrightarrow k'$, $\forall k,k'$ are induced).
In the opposite case of $l=N$ (just one bath for the whole system) no transition between different $k$
is allowed. In the intermediate case each mode interacts with the other modes in an interval of width $2\pi l/N$.

It is important to notice  that correlations between baths over a finite distance would not
change qualitatively our picture as long as we focus on the critical properties of the model.
Indeed, near criticality the divergence of the correlation length makes the details of the bath
correlations over microscopic distances unimportant.
For the same reason, as long as one is interested in the low-$T$ properties of the bath, the
specific value of $l$ is not relevant provided $l/N \rightarrow 0$ in the thermodynamic limit.
Specifically, for all values of $l$ such that $T\ll\left(\frac{1}{l}\right)^{z}$
the same dissipative dynamics is obtained, since transitions with large $\Delta k$
(and hence large energy) are thermally suppressed (we used $E\propto k^{z}$ at a fixed $T$).
In this regime, therefore, the system cannot resolve the microscopic details of different system-bath
couplings (i.e., whether $l=1, 2, 3,\dots$).
In the following, we thus focus on the case $l=1$: specific high-temperature and non-critical behaviors
for different $l$ could be easily investigated within the same scheme considered below.
Only if $l = N$ the dynamics of the system changes qualitatively.

\section{Kinetic equation} \label{sec:Kinetic-equation}

In this section we derive a kinetic equation for the Green's function of the Jordan-Wigner fermions
within the Keldysh formalism. In terms of this Green's function we will then calculate both the excitation
and the energy densities, Eq.~(\ref{eq: excitations density}) and (\ref{eq:eenergy density}).
Our analysis in terms of a kinetic equation will provide support for the scaling laws obtained above,
Eqs.~(\ref{eq:Ebath_generic_QPT})-(\ref{eq:ETAbath_genericQPT}), while allowing us to study the non-universal
dynamics beyond the limit of applicability of the scaling approach.
%

The fermionic Keldysh Green's function is a matrix in Nambu space defined by
\begin{equation}
[G_{k}(t_{1},t_{2})]_{i,j} \equiv -i \left\langle \mathcal{T}_{\gamma} \,
\Psi_{ki}^{\phantom{\dagger}}(t_{1}) \Psi_{kj}^{\dagger}(t_{2})\right\rangle
\label{eq: Keldysh GF}
\end{equation}
(see appendix A1 for notations) where $\gamma$ is the Keldysh contour. In the following we will neglect
the initial correlations between system and bath \cite{Rammer86}.
Hence $\gamma$ consists of just a forward and backward branch on the real time axis.
Below we will sketch of the main steps of the derivation: the remaining details can be found in Appendix A1-A2.

The starting point of our derivation is the Dyson's equation in its integro-differential form
\begin{eqnarray}
\left[ i \partial_{t_{1}}-\mathcal{\hat{H}}_{k}(t_{1})\right] G_{k}(t_{1},t_{2}) &=&
\delta(t_{1}-t_{2}) \label{eq: Dyson eq}\\
 &+& \int_{\gamma} \! d\bar{t} \, \Sigma_{k}(t_{1},\bar{t}) \, G_{k}(\bar{t},t_{2}) \nonumber
\end{eqnarray}
and an analogous one obtained by differentiation with respect $t_{2}$. Here $\Sigma_k$ is the
self-energy associated to the interaction of the system with the bath.
In order to compute the energy and excitations density we
need to find the equal-time statistical Green's functions.
The latter are defined as
\begin{eqnarray}
[G_{k}^{<}(t_{1},t_{2})]_{i,j} &\doteq& i\langle\Psi_{k,j}^{\dagger}(t_2)\Psi_{k,i}^{\phantom{\dagger}}(t_1)\rangle
\label{eq:lesser GF}\\
{}[G_{k}^{>}(t_{1},t_{2})]_{i,j} &\doteq& -i\langle\Psi_{ki}^{\phantom{\dagger}}(t_{1})\Psi_{kj}^{\dagger}(t_{2})\rangle
\label{eq: greater GF}
\end{eqnarray}
An equation for these correlators can be obtained from Eq.~(\ref{eq: Dyson eq}) by using standard
techniques~\cite{HaugBOOK} (see Appendix A1 for details). For the equal-time Green's function
$G_{k}^{<}(t,t)$ we obtain:
\begin{eqnarray}
i \partial_{t}G_{k}^{<} &=& \left[\mathcal{\hat{H}}_{k}, G_{k}^{<}\right] +
\label{eq: Kinetic lesser greater}\\
 &  & \Sigma_{k}^{>}\cdot G_{k}^{<} - \Sigma_{k}^{<}\cdot G_{k}^{>} + G_{k}^{<} \cdot \Sigma_{k}^{>}
      - G_{k}^{>} \cdot\Sigma_{k}^{<} \nonumber \;,
\end{eqnarray}
where the dots indicate the convolution:
$$
\Sigma_{k}^{>} \cdot G_{k}^{<} \doteq \int_{0}^{t}d\bar{t}\,\Sigma_{k}^{>}(t,\bar{t}) G_{k}^{<}(\bar{t},t) \; .
$$

\begin{figure}[h]
\includegraphics[scale=0.28]{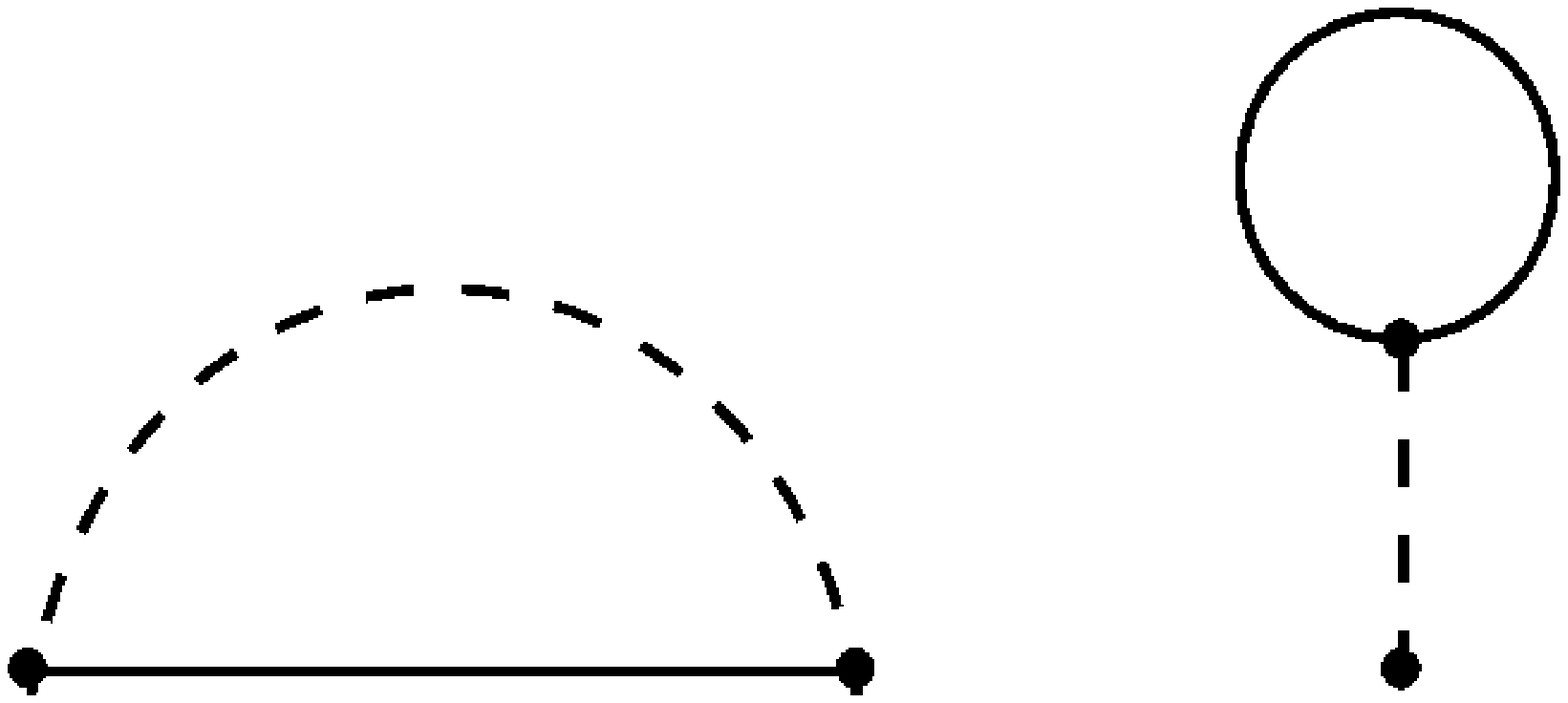}\put(-135,0){{\bf a)}}\put(-35,0){{\bf b)}}
\caption{\label{fig: diagram}
Lowest-order diagrams contributing to the self-consistent Born approximation:
dashed  lines correspond to the non-interacting bath Green's function $g$,
while solid lines to the interacting system Green function $G$.
{\bf a)} corresponds to Eq.~(\ref{eq:SIGMA_keldysh}), while {\bf b)} to Eq.~(\ref{eq:SIGMA polaronic}). }
\end{figure}

In order to proceed with the solution of  Eq.~(\ref{eq: Kinetic lesser greater}) it is now important to discuss the approximations
we make for the self-energy. Let us first notice that long-time correlations induced by the bath
may change the universality class of the transition, by renormalizing
the low energy spectrum of the system \cite{Werner07}.
As previously mentioned, we will not consider this case here.
Therefore, we assume that the bosons have a non-zero inverse lifetime
$\Gamma\ll T$ which provides a natural cutoff-time for the bath correlation functions.
Within this assumption it is now possible to describe the kinetics of the system
using a Markov approximation together with a self-consistent Born approximation.
The latter is justified for weak system/bath coupling ($\alpha \ll 1$) and is
represented diagrammatically in Fig.~\ref{fig: diagram}-(a).
%
%
We will neglect the tadpole diagram (b), which represents just a small shift of the energy levels.

By going  to the interaction picture
\begin{eqnarray*}
\tilde{G}_{k}(t_{1},t_{2}) &\doteq& \mathcal{\hat{U}}_{k}^{\dagger}(t_{1}) G_{k}(t_{1},t_{2})
\hat{\mathcal{U}}_{k}^{\phantom{\dagger}}(t_{2}) \;,
\end{eqnarray*}
it is now evident that, within our assumptions, the evolution of $\tilde{G}_{k}$
can be considered slow as compared to that of the bath correlators appearing in the self-energies.
Using this separation of time scales it is possible to implement
the Markov approximation and transform the general integro-differential kinetic equation into
a simple differential equation (see Appendix A1-A2).
%
%
We then obtain, in the case in which each spin is coupled to its own bath ($l=1$), the kinetic equation
%
%
\begin{eqnarray}
\partial_{t}G_{k}^{<} &+& i\left[\mathcal{H}_{k},\, G_{k}^{<}\right] = \label{eq:Kinetic}\\
\frac{1}{N} \sum_{q} \tau^{z} ({\bf 1}+iG_{q}^{<}) \hat{D}_{qk} G_{k}^{<} &+&
\tau^{z} G_{q}^{<} \hat{D}_{kq}^{\dagger} ({\bf 1}+iG_{k}^{<}) + {\it H.c.} \nonumber
\end{eqnarray}
where
\begin{equation}
\hat{D}_{qk}=i\int_{0}^{\infty}\! ds\, g^{>}(s) \, {\hat{\mathcal{U}}}_{q}^{\dagger}(t,t-s) \, \hat{\tau}^{z}
\, {\hat{\mathcal{U}}}_{k}(t,t-s) \;,
\label{eq: D matrices main}
\end{equation}
$g^{>}(t)=-i\left\langle X_{q}(t)X_{q}(0)\right\rangle$, and ${\hat{\mathcal{U}}}_{k}(t_{0},t)$ is
the evolution operator satisfying
$i\partial_{t}{\mathcal{\hat{U}}}_{k}={\hat{\mathcal{H}}}_{k}(t){\hat{\mathcal{U}}}_{k}$.
The left-hand-side of Eq.~(\ref{eq:Kinetic}) represents the free evolution term,
while the right-hand-side describes the scattering between the $k$ modes mediated by the bath degrees of freedom.
Notice that the number of equations scales linearly with the system size $N$, in contrast
to conventional systems of master equations whose number scales exponentially with $N$ as a result of
the fact that the full density matrix (i.e. all $m-$points Green functions) is considered.
The fact that we are considering only the two-point Green's function
self consistently using the Born approximation is responsible for the
non-linear nature of Eq.~(\ref{eq:Kinetic}), in contrast to the linearity
of the master equation.

In the eigenbasis of the Hamiltonian $\mathcal{\hat{H}}_{k}$ the Green's function can be parameterized as
\begin{equation}
-iG_{k}^{<}=\left(\begin{array}{cc}
\mathcal{P}_{k} & \mathcal{C}_{k}\\
\mathcal{C}_{k}^{*} & 1-\mathcal{P}_{k}\end{array}\right)
\label{eq: rho H basis}
\end{equation}
where $\mathcal{P}_{k}=\langle\eta_{k}^{\dagger}\eta_{k}\rangle$ is the population of the excited mode $k$
and $\mathcal{C}_{k}=\left\langle \eta_{-k}\eta_{k}\right\rangle$ can be regarded as a ``coherence''  term~\cite{CohenBOOK}.
In the static case, where the evolution operator is $\hat{\mathcal{U}}_{k}=\exp(-i\mathcal{\hat{H}}_{k}t)$,
the stationary solution of the kinetic equation (\ref{eq:Kinetic}) is correctly the thermal equilibrium one:
$\mathcal{C}_{k}=\mathcal{C}_{k}^{th}=0$ and $\mathcal{P}_{k}=\mathcal{P}_{k}^{th}=(e^{\Lambda_{k}/k_{B}T}+1)^{-1}$
(the Fermi function).

Finally, once the solution of the kinetic equation (\ref{eq:Kinetic}) is obtained, the density of excitations
and energy produced during the quench can be expressed as
\begin{eqnarray}
\mathcal{E} &=& \frac{1}{N}\sum_{k>0}\mathcal{P}_{k}
\label{eq:excitation_density}\\
{E} &=& \frac{1}{N}\sum_{k>0}\Lambda_{k}\mathcal{P}_{k}
\label{eq:energy density}
\end{eqnarray}
%
%
We conclude this section by commenting on a useful approximation to evaluate numerically
the kernel of $\hat{D}_{qk}$, Eq.~(\ref{eq: D matrices main}), discussed in full detail in Appendix A2.
It consists in approximating the evolution operator ${\hat{\mathcal{U}}}_{k}$ appearing in $\hat{D}_{qk}$
with ${\hat{\mathcal{U}}}_{k}(t,t-s)\simeq\exp\left( i\hat{\mathcal{H}}_{k}(t)s\right)$,
thus obtaining
\begin{equation}
\hat{D}_{qk}\simeq i\int_{0}^{\infty}\! ds\, g^{>}(s)\exp\left( -i\hat{\mathcal{H}}_{k}(t)s\right)
\hat{\tau}^{z}\exp\left( i\hat{\mathcal{H}}_{k}(t)s\right) \;.
\label{eq: Dapprox main}
\end{equation}
This is again consistent with the separation of time scales mentioned above, and in particular with the Markov
approximation. Indeed, while the exact relaxation rate matrix (\ref{eq: D matrices main})
depends on the velocity of the quench, if the quench is slow on the time scale characteristic
of the bath, the correlation function $g^{>}(s)$ can be seen as strongly peaked at $s=0$.
Hence the system can be considered ``frozen'' at the instantaneous value of $h(t)$ and, consistently,
its evolution operator is the exponential of the Hamiltonian.

\section{Relaxation time} \label{sec:Relaxation-time}

In order to make further progress in understanding the quench dynamics of the system
we will first extract from the kinetic equation the characteristic
relaxation time for the populations of the excitations $\mathcal{P}_{k}$
(see Eq.(\ref{eq: rho H basis})) as a function of the magnetic field $h$ and the temperature $T$.
For this purpose, it is sufficient to consider only the diagonal elements of Eq.~(\ref{eq: rho H basis}).
This is equivalent to the so called ``secular approximation'' for the master equation \cite{CohenBOOK},
which is valid for weak couplings ($\alpha\ll 1$ in the present case).
For generic $N$, we deal with a set of ${N}/{2}$ equations (only ${N}/{2}$ modes are independent)
of the form:
\[
\frac{d}{dt}\mathcal{P}_{k} = a_{k} + \sum_{q} b_{kq} \mathcal{P}_{q}
                                    + \sum_{q} c_{kq} \mathcal{P}_{k} \mathcal{P}_{q} \;.
\]
The asymptotic relaxation can be studied by linearizing the previous set of equations near the thermal
equilibrium fixed point.
We obtain, with the vector notation $\delta\mathcal{\underline{P}}=\left(\delta\mathcal{P}_{1},
\delta\mathcal{P}_{2},\dots,\delta\mathcal{P}_{N/2}\right)^{\rm tr}$ where
$\delta\mathcal{P}_{k}=\mathcal{P}_{k}-\mathcal{P}_{k}^{th}$:
\begin{equation}
\frac{d}{dt}\delta\mathcal{\underline{P}}=-\mathcal{R\ } \delta\mathcal{\underline{P}}
\label{eq: diffeq R}
\end{equation}
where non-linear terms in $\delta\mathcal{P}_{k}$ have been neglected.
The diagonal and off-diagonal elements of the $N/2\times N/2$ matrix $\mathcal{R}$ are:
\begin{widetext}
\begin{eqnarray}
\mathcal{R}_{kk} &=& \frac{2}{N}\sum_{q>0,q\ne k}
\left[ \mathcal{G}_{q}^{th} \left(1-\cos(\theta_{k}+\theta_{q})\right) g[-\Lambda_{k}-\Lambda_{q}]
     + \mathcal{G}_{q}^{th} \left(1+\cos(\theta_{k}+\theta_{q})\right) g[\Lambda_{k}-\Lambda_{q}] \right.
\label{eq: Adiag}\\
 && \hspace{15mm} \left. + \mathcal{P}_{q}^{th} \left(1-\cos(\theta_{k}+\theta_{q})\right) g[\Lambda_{k}+\Lambda_{q}]
          + \mathcal{P}_{q}^{th} \left(1+\cos(\theta_{k}+\theta_{q})\right) g[-\Lambda_{k}+\Lambda_{q}]
   \right]\nonumber \\
 &+& \frac{2}{N}\ 4\sin^{2}\theta_{k} \left(\mathcal{G}_{k}^{th} g[-2\Lambda_{k}]
                                           +\mathcal{P}_{k}^{th} g[2\Lambda_{k}]\right)\nonumber
\end{eqnarray}
\begin{eqnarray}
\mathcal{R}_{kq} &=& \frac{2}{N}\left[-\mathcal{G}_{k}^{th}\left(1+\cos(\theta_{k}+\theta_{q})\right)
g[-\Lambda_{k}+\Lambda_{q}]+\mathcal{G}_{q}^{th}\left(1-\cos(\theta_{k}+\theta_{q})\right)
g[-\Lambda_{k}-\Lambda_{q}]\right.
\label{eq:Aoffdiag}\\
 &  & \hspace{5mm} \left.-\mathcal{P}_{q}^{th}\left(1+\cos(\theta_{k}+\theta_{q})\right)
g[\Lambda_{k}-\Lambda_{q}]+\mathcal{P}_{q}^{th}\left(1-\cos(\theta_{k}+\theta_{q})\right)
g[\Lambda_{k}+\Lambda_{q}]\right]\nonumber
\end{eqnarray}
\end{widetext}
where $g[E]$ is the Laplace transform of bath correlator (\ref{eq:g[E]}), $\mathcal{G}_{k}^{th}$ is the
thermal equilibrium value of the population of the ground-state of mode $k$,
$\mathcal{G}_{k}^{th}=1-\mathcal{P}_{k}^{th}$, and
\begin{equation}
\theta_{k}=\arccos -\frac{(\cos k+h)}{\Lambda_{k}} \;.
\label{eq: theta_k}
\end{equation}
The eigenvalues of $\mathcal{R}$, $\{\lambda_{i}\}$, are the characteristic
relaxation rates of the long-time dynamics. Hence the solution of Eq.~(\ref{eq: diffeq R})
for each population would be a linear combination containing all the
characteristic relaxation times:
\[
\delta\mathcal{P}_{k}=\sum_{j}r_{kj}e^{-\lambda_{j}t}
\]
At long times $t\gg\left(\min_{j}\lambda_{j}\right)^{-1}\doteq\tau$
all modes relax with the same relaxation time $\tau$.
In the following we first analyze the longest relaxation time $\tau$, extending the
results presented in Ref.~\onlinecite{PatanePRL}; we then study the structure
of the entire spectrum of relaxation times.
%
\begin{figure}[h]
\includegraphics[scale=0.65]{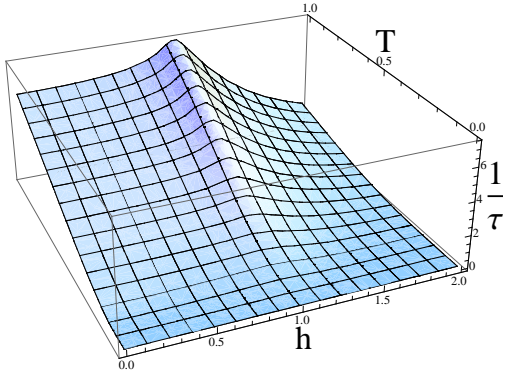}
\caption{\noun{\label{fig: Relaxation time 3D} } Relaxation rate $1/\tau$
as a function of $T$ and $h$ for $N=400$ (here $\gamma=1$,and $s=1$). }
\end{figure}

In Fig.~\ref{fig: Relaxation time 3D} we show the general behavior of $\tau$ in the finite
temperature phase-diagram, calculated by numerically diagonalizing the matrix $\mathcal{R}$.
As $T\rightarrow 0$, $\tau$ diverges and close to the critical point two different behaviors are
found in the semiclassical regions and in the quantum critical region (see also Fig.\ref{fig:fig1}):
\begin{equation}
\tau^{-1}\propto  \begin{cases}
                   T^{1+s} & T\gg\Delta\\
                   e^{-\Delta/T} & T\ll\Delta
                   \end{cases}
\label{eq: tau energy}
\end{equation}
These relations extend the results obtained for the relaxation rate in Ref.~\onlinecite{PatanePRL}
to the generic case of non-Ohmic baths and give the exponent $\theta$ as $\theta=1+s$.

\begin{figure}[h]
\includegraphics[scale=0.8]{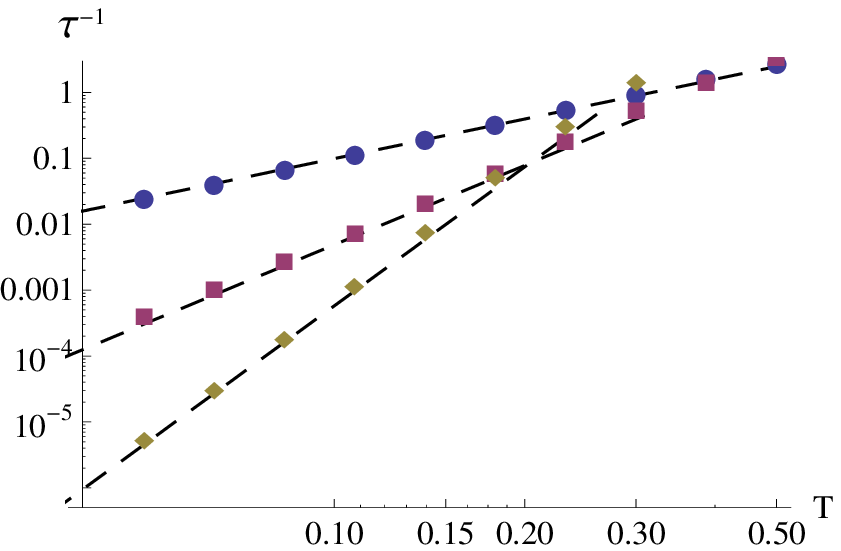}\put(-165,90){$s=1$}\put(-155,65){$s=3$}\put(-125,35){$s=6$}\\
\includegraphics[scale=0.8]{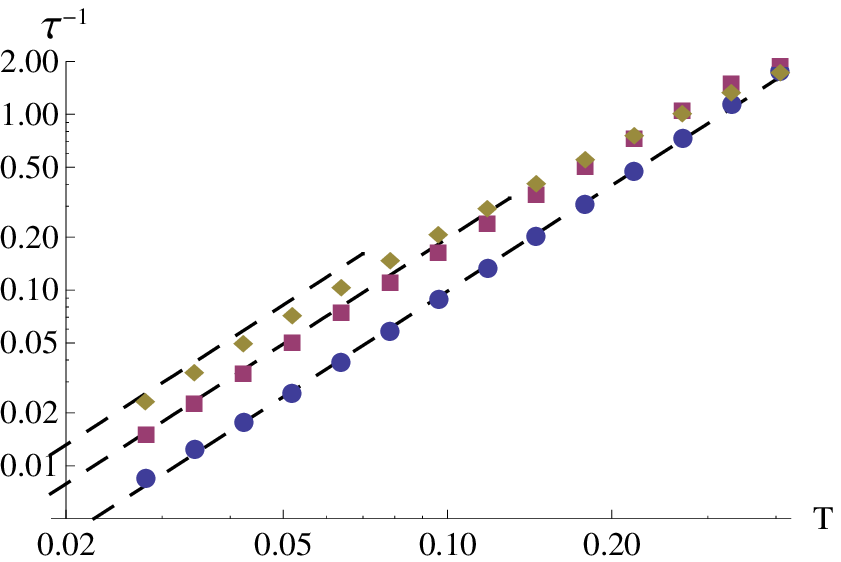}
\caption{\noun{\label{fig: tau s-gamma} } Relaxation rate $1/\tau$ as a function
of $T$ as obtained from the exact diagonalization of $\mathcal{R}$
(symbols) and the approximation in Eq.~(\ref{eq: tau diagonal}) (dashed lines).
Upper panel: $s=1$, $3$, $6$, with $\gamma=1$;
lower panel: $s=1$ with different $\gamma=0.3$, $0.5$, $1$ from top to bottom. }
\end{figure}

An analytic expression for the power-law scaling inside the quantum critical region can be obtained
by approximating the smallest eigenvalue of $\mathcal{R}$ with the smallest diagonal element.
This is justified by the fact that the off-diagonal elements are of the order $O(1/N)$ (see Eq.~(\ref{eq:Aoffdiag})).
For $h=h_{c}=1$, considering the gapless mode $k=\pi$, we have from Eq.~(\ref{eq: Adiag}) in the continuum limit:
\begin{eqnarray}
\tau_{diag}^{-1} &\doteq\mathcal{R}_{\pi\pi}& =
\frac{2}{\pi} \int_{0}^{\pi} \! dq\,
\left( \mathcal{G}_{q}^{th} g[-\Lambda_{q}] + \mathcal{P}_{q}^{th} g[\Lambda_{q}] \right) \nonumber \\
 &  & =4\alpha\int_{0}^{\pi}\! dq \, \frac{\Lambda_{q}^{s}}{\sinh\left( \Lambda_{q}/T\right) }\nonumber \\
 &  & \hspace{-2cm}\simeq 8\alpha \, (1-2^{-1-s}) \, \Gamma(1+s) \, \zeta(1+s)\left(\gamma/T\right)^{-1-s}
\label{eq: tau diagonal}
\end{eqnarray}
where $\Gamma$ and $\zeta$ are the Gamma and the zeta functions,
and we used the critical dispersion relation $\Lambda_{q}\simeq\gamma(\pi-q)$
obtained by linearizing Eq.~(\ref{eq:energy spectrum}) around the gapless point $k=\pi$
(we extended the integration to $-\infty$ since at low-temperature only the low-energy
modes contribute to the integral).
Fig.~\ref{fig: tau s-gamma} demonstrates that the analytical expression in
Eq.~(\ref{eq: tau diagonal}) agrees very well with the numerical solution
(obtained by diagonalizing $\mathcal{R}$), especially at low temperature.
%

As we have shown in Fig.~\ref{fig: tau s-gamma}, inside the quantum
critical region the exponent $\theta$ is universal within the range
of anisotropy $0<\gamma\le 1$ where the system belongs to the Ising
universality class. This suggests a relation between $\theta$ and
the critical indexes of the quantum phase transition. Indeed, it can
be shown, within the Fermi golden rule (see Appendix B), that for a
generic system coupled to a bosonic bath the following expression
holds inside the quantum critical region:
\begin{equation}
\tau^{-1}\propto T^{s+d/z} \;.
\label{eq:tau general}
\end{equation}

\begin{figure}[h]
\includegraphics[scale=0.65]{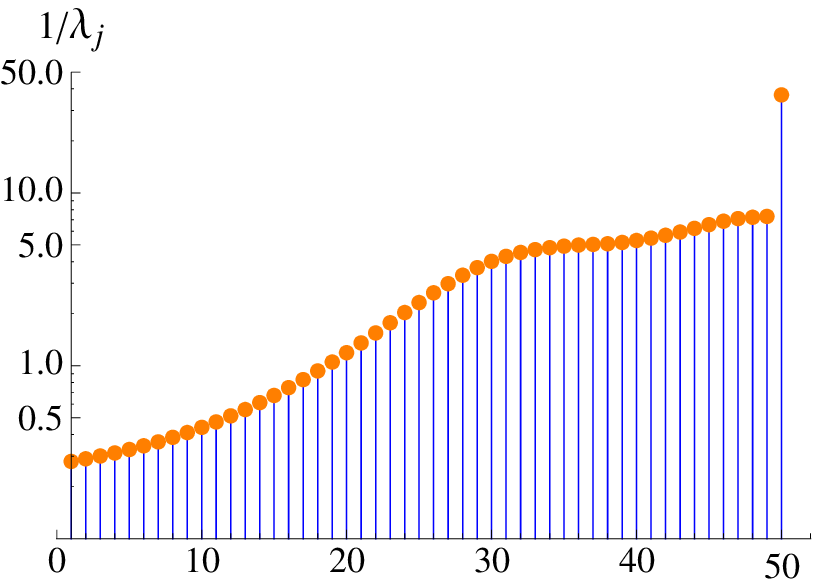}\put(-100,80){$h=0.8$}\\
\includegraphics[scale=0.65]{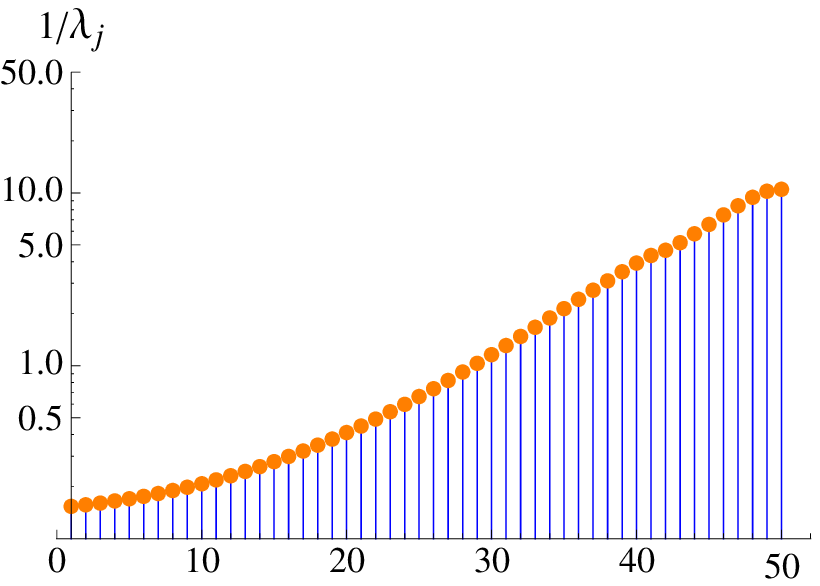}\put(-100,70){$h=1$}
\caption{\noun{\label{fig:Eigenvalues of A}} Spectrum of the eigenvalues of
$\mathcal{R}$ for the Ising model ($\gamma=1$) with Ohmic baths
($s=1$), for $N=100$. The values of temperature $T=0.1$ and magnetic
field $h=0.8$, or $1$ are chosen to belong, respectively, to the
semiclassical and the quantum critical region. }
\end{figure}

\begin{figure}
\includegraphics[scale=0.75]{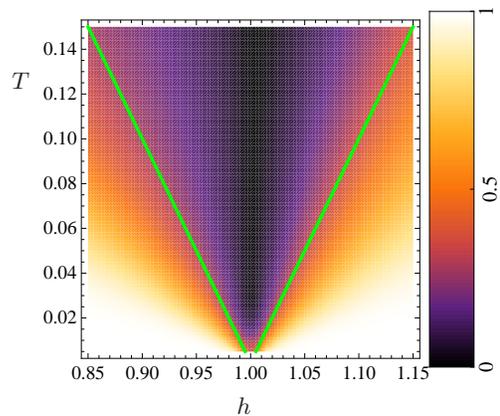}\put(-185,120){$T$}\put(-100,-3){$h$}\label{fig: tau_gap}
\caption{Ising model ($\gamma=1$) with Ohmic baths ($s=1$). Relative gap
between the two longest relaxation times: $(\lambda_{2}^{-1}-\lambda_{1}^{-1})/\lambda_{1}^{-1}$
($\lambda_{1,2}$ being the two lowest eigenvalue of $\mathcal{R})$;
crossover lines $T=T_{cross}=|h-1|$ are plotted for comparison. }
\end{figure}

An important feature of the relaxation dynamics can be extracted by
analyzing the spectrum of the eigenvalues $\{\lambda_{j}\}$ of $\mathcal{R}$.
In Fig.~\ref{fig:Eigenvalues of A} the $\lambda_{j}$'s are shown for
some values of temperature and magnetic field. We find that in the
semiclassical regions $T\ll\Delta$ the smallest eigenvalue of $\mathcal{R}$
is separated from the rest of the spectrum by a gap (even in the $N\rightarrow\infty$ limit).
On the contrary, inside the quantum critical region such eigenvalue
merges with the rest of the spectrum. This can be quantified by the
relative gap of the spectrum of relaxation times, that is identified
by $(\lambda_{2}^{-1}-\lambda_{1}^{-1})/\lambda_{1}^{-1}$, being $\lambda_{1,2}$ the lowest
eigenvalues of $\mathcal{R}$ (see Fig.~\ref{fig: tau_gap}).
Such result indicates that while the exponential divergence of the relaxation time
$\tau\propto\exp\left\{ \Delta/T\right\} $ in the semiclassical regions is due to an isolated eigenvalue,
the long-time behavior in the quantum critical region is, instead, built up by a continuum of eigenvalues
contributing to the $\tau^{-1}\propto T^{2}$ scaling.
%

\section{Adiabatic quenches} \label{sec:Adiabatic-quenches}

Equipped with the kinetic equation and the knowledge of the scaling of the relaxation times,
 we now analyze the quench dynamics of the model in Eq.~(\ref{eq:H_TOT}))
by solving the kinetic equation (\ref{eq:Kinetic}) numerically.
The system is initialized at $h_{i}\gg h_{c}$
in equilibrium with the bath at a fixed temperature $T\ll\Delta(h_{i})$, and the transverse
field is then ramped linearly $h(t)=h_{i}-vt$ down to a final value $h_{f}$
(the bath temperature is kept fixed).

\begin{figure}
\includegraphics[scale=0.35]{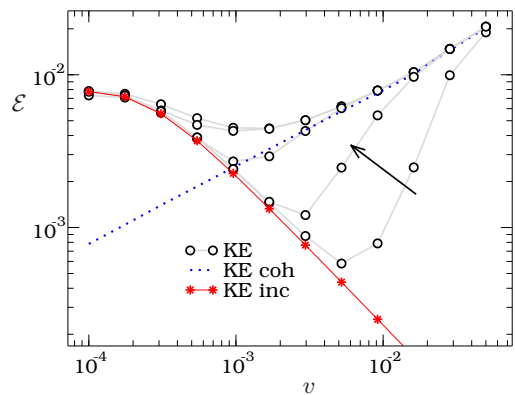}\put(-190,100){$\mathcal{E}$}\put(-80,-7){$v$}
\caption{\label{fig:finite_size}  Density of excitation $\mathcal{E}$ (circles) Vs quench velocity for different system sizes
$N=26,~50,~100,~200,~400,~800$ from bottom to top, according to the arrow; the points corresponding to $N=800$ and $N=400$
are indistinguishable. Parameters are set $\alpha=0.01$, $T=0.1$ $\gamma=1$ and $s=1$ and the quench is halted at $h_f=0.8$.
Dotted line is the coherent contribution $\mathcal{E}_{coh}$ obtained for $\alpha=0$
and stars represent the incoherent contribution $\mathcal{E}_{inc}$ due to thermal excitation (see text);
both curves refere to $N=800$, even if for $\mathcal{E}_{inc}$ the same curve is obtained already at $N\sim 30$.
 }
\end{figure}

In  Fig. \ref{fig:finite_size} we plot the density of excitations as a function of the quench velocity
 for different system sizes (a similar behavior is obtained for the density of energy).
Additionally we considered separately the coherent ($\mathcal{E}_{coh}$)
 and incoherent contribution ($\mathcal{E}_{inc}$) to the final density of excitations.
The first one is obtained by integrating the kinetic equation for $\alpha=0$, i.e. no coupling with the bath.
The incoherent term is due to thermal excitations created by the bath and it is obtained by integrating the kinetic equation
and  ignoring the unitary evolution term $i[\hat{\mathcal{H}}_k,\hat{G}^<_k]$,  responsible for the coherent excitation process.

\begin{figure}
\includegraphics[scale=0.35]{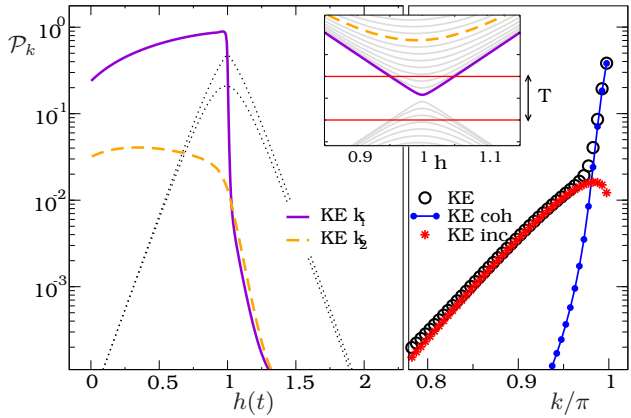}\put(-235,128){$\mathcal{P}_{k}$}\put(-150,-7){$h(t)$}\put(-30,-7){$k/\pi$}
\caption{\label{fig:p_k} Populations of the excited states $\mathcal{P}_k$ for $N=400$, $v=0.0017$
and the same values of Fig. \ref{fig:finite_size}.
Left: dynamics of two low energy modes $k_{1}\sim\pi$ (dashed line) and
$k_{2}\sim0.9\pi$ (dotted-dashed line) obtained  by solving the kinetic equation
 (instantaneous thermal equilibrium values are plotted as reference as dotted lines).
The Inset shows  the energy levels near the critical point and the scale of temperature;
 marked levels of the excited band refer to $k_{1}$ and $k_{2}$. The energy gap closes at $k=\pi$.
 Right: distribution of $\mathcal{P}_{k}$ as a function of the mode k at $h(t)=1$.
Stars represent the  excitations created incoherently and
 triangles are the coherent excitations produced in the case of no coupling to the bath.
The two excitation mechanisms act on different energy scales,
 being lowest energy modes coherently  populated and the highest one thermally excited. }
\end{figure}

In order to understand the two excitation mechanisms we analyse directly the dynamics of the populations $\mathcal{P}_k$.
From the results shown in  Fig \ref{fig:p_k} (left) it emerges that excitations are generated close to the critical point
and when the system is driven in the semiclassical region (and $T\ll\Delta$) they are relaxed out by the bath.
The density of excitation generated is the sum of the the incoherent and coherent contribution,
thus proving the validity of the {\it Ansazt}  (\ref{eq:ebath+eclean}) and (\ref{eq:etabath+etaclean}) (see  Fig. \ref{fig:p_k} (right)).

For the XY model the integration of Eqs.~(\ref{eq: Einc int}) and (\ref{eq:ETAinc int}) can be performed
explicitly by using the critical spectrum $\Lambda_{k}\sim\gamma(\pi-k)$.
We obtain:
\begin{eqnarray}
\mathcal{E}_{inc}& = & \frac{\log2}{2\pi\gamma}T\left(1-e^{-2T/(\tau v)}\right) \label{eq:e BATH XY}\\
{E}_{inc} & = & \frac{\pi}{24\gamma}T^{2}\left(1-e^{-T/(\tau v)}\right) \label{eq:etaBATH XY}
\end{eqnarray}
where the latter holds for quenches halted at $h_{f}=h_{c}$. In the previous formulas, the expression derived
for $\tau$ in Eq.~(\ref{eq: tau diagonal}) can be used to get a fully analytical expression.
Expanding the exponentials in Eqs.~(\ref{eq:e BATH XY}) and (\ref{eq:etaBATH XY}) we obtain
\begin{eqnarray}
\mathcal{E}_{inc} & \simeq & \frac{8\log 2}{\pi}~\varphi(s)~\alpha\gamma^{-2-s}~v^{-1}T^{3+s} \label{eq:e BATH XY powerlaw}\\
{E}_{inc} & \simeq  & \frac{\pi}{3}~\varphi(s)~\alpha\gamma^{-2-s}~v^{-1}T^{4+s} \label{eq:etaBATH XY powerlaw}
\end{eqnarray}
where $\varphi(s)=(1-2^{-1-s})\Gamma(1+s)\zeta(1+s)$. The previous relations are
 consistent with Eqs.(\ref{eq:Ebath_generic_QPT}) and (\ref{eq:ETAbath_genericQPT})
with $\theta=1+s$ (see Eq.~(\ref{eq:tau general})).
%
\begin{figure}
\includegraphics[scale=0.4]{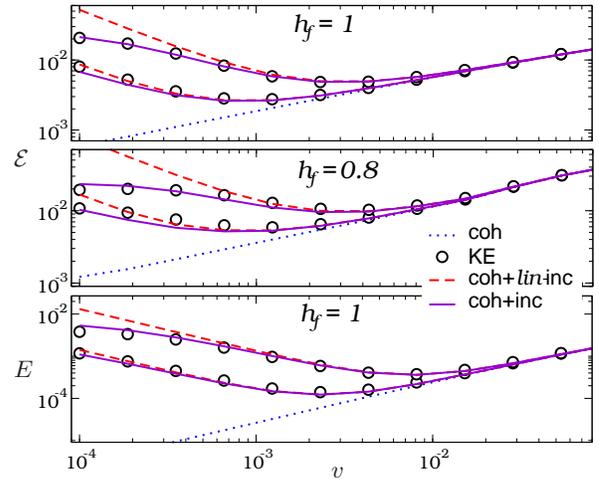}\put(-220,115){$\mathcal{E}$}\put(-220,35){${E}$}\put(-100,-3){$v$}
\caption{\label{fig:E and Eta} Density of energy (lowest panel) and of excitations
versus quench velocity $v$ for $h_{f}=0.8$, and $1$. Parameters are set to $\gamma=0.7$,
$s=1.5$, $\alpha=0.01$, and $T=0.15$, or $0.1$ (upper and lower curves of each panel).
Circles are obtained by solving the kinetic equation;
dotted lines are the coherent contributions $\mathcal{E}_{coh}$ and
${E}_{coh}$ evaluated by solving the kinetic equation for $\alpha=0$;
a fit gives correctly $\mathcal{E}_{coh}\propto\sqrt{v}$
and ${E}_{coh}\propto v$ consistently with the KZ scaling-law for excitations (\ref{eq: E KZ})
and the modified scaling we derived for the energy density (\ref{eq: eta KZ}).
Solid and dashed lines are Eqs.~(\ref{eq:ebath+eclean}) and (\ref{eq:etabath+etaclean})
using for the incoherent contributions the expressions (\ref{eq:e BATH XY}) and (\ref{eq:etaBATH XY})
and their linearized forms (\ref{eq:e BATH XY powerlaw}) and (\ref{eq:etaBATH XY powerlaw}),
respectively.}
\end{figure}
\begin{figure}
\includegraphics[scale=0.4]{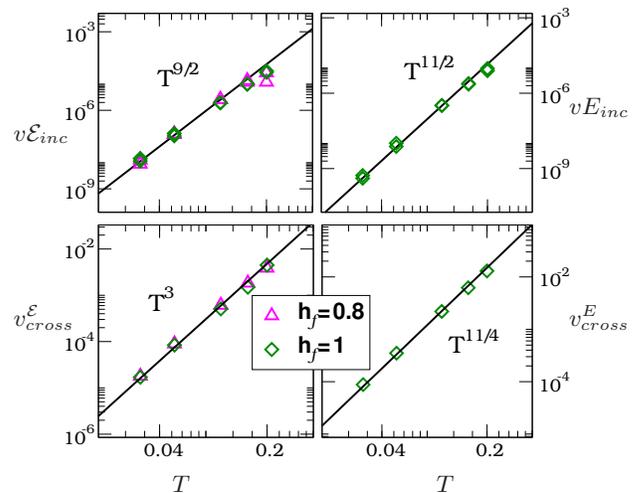}\put(0,130){$vE_{inc}$}\put(0,50){$v^{E}_{cross}$}\put(-210,120){$v\mathcal{E}_{inc}$}\put(-210,50){$v^{\mathcal{E}}_{cross}$}\put(-50,-13){$T$}\put(-150,-13){$T$}
\caption{\label{fig:v_cross_INC}  System and bath parameters are fixed as in of Fig. \ref{fig:E and Eta}.
Upper panel: Data collapse of $\mathcal{E}_{inc}$ and $E_{inc}$
obtained from the kinetic equation; data refer
 to $10^{-3}\lesssim v \lesssim 10^{-2}$ (data relative to  $h_f=1$ for  $\mathcal{E}_{inc}$
are rescaled by a factor $2$, since in this case only half quantum critical region is crossed).
Lower panel:  scaling of $v_{cross}$ is obtained
equating $\mathcal{E}_{inc}=\mathcal{E}_{coh}$ and analogously for $E$.
The fits confirm the scaling predicted by (\ref{eq:Ebath_generic_QPT}), (\ref{eq:ETAbath_genericQPT})
and (\ref{eq:E vbath_generic_QPT}), (\ref{eq:ETA vbath_generic_QPT}), that, for the specific case
of $s=1.5$ considered, are shown in their corresponding plots.}
\end{figure}

\begin{figure}
\includegraphics[scale=0.4]{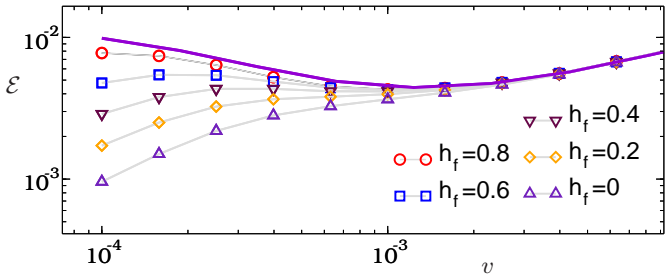}\put(-250,65){$\mathcal{E}$}\put(-70,-3){$v$}
\caption{\label{fig:E hf}Density of excitations versus quench velocity $v$ for
$h_{f}=0.8,\ 0.6,\ 0.4,\ 0.2,\ 0$. Parameters are set to $\gamma=1$,
$s=1$, $\alpha=0.01$ and $T=0.1$. Upper solid line is the scaling-law (\ref{eq:ebath+eclean})
using the expression (\ref{eq:e BATH XY}).
Decreasing the value of $h_{f}$, the crossing of the semiclassical
region, after the critical point, becomes more relevant at low $v$ and
scaling no longer holds strictly. }
\end{figure}

In Fig.~\ref{fig:E and Eta} the density of excitations and energy
obtained from the solution of kinetic equation is compared with the
scaling-law derived in Sec.~\ref{sec:Scaling-analysis} using the
specific expressions, Eqs.~(\ref{eq:e BATH XY}) and (\ref{eq:etaBATH XY}),
derived above for the XY model. The scaling-laws are found in good
agreement with the numerical data.
The results shown in Fig. \ref{fig:v_cross_INC} further confirm the scaling as a function
of the temperature and the relations  (\ref{eq:E vbath_generic_QPT}), (\ref{eq:ETA vbath_generic_QPT})
 for the crossover velocity.

 Finally we comment on the role of the final value of magnetic field $h_{f}$ at which the
quench is halted.  The agreement with the scaling {\it Ansatz} becomes worse for decreasing $h_{f}$
 (see Fig.~\ref{fig:E hf}). This is due
to the non-critical relaxation induced by the bath when the system
crosses the semiclassical region after the critical point
(see Fig.~\ref{fig:fig1} and Fig.~\ref{fig:p_k} left). At low $v$ the time spent therein at a relative
low-temperature $T\ll\Delta$ is so long that the bath is able to
relax the excitations created close to the critical point.

\section{Conclusions}
\label{conclusions}

We have studied the dynamics of a quantum critical system coupled
to a thermal reservoir and subject to an adiabatic quench
across its quantum critical point. We considered the regime of weak
coupling, low-temperature and slow quench velocity.

The bath has two effects on the system: the first one is to create
excitations inside the quantum critical region and the second one
is to trigger the relaxation of the excitations created close to the critical point
when the system is driven in the semiclassical region (see Fig.~\ref{fig:fig1}).
While the first mechanism is universal, being entirely ruled by the
critical properties of the low-energy spectrum, the latter depends
on the details of the system far-off the critical point. Hence, as
far as the evolution is halted close to the critical point and the
non-critical relaxation mechanism is negligible, universal scaling
behavior is recovered. We derived scaling-laws for the density of
energy produced by the quench at finite temperature extending the
previous results obtained for the density of excitations in Ref.~\onlinecite{PatanePRL}.

To check the validity of the scaling-laws, we considered the specific
case of the quantum XY model (\ref{eq: H_S}) coupled locally to a
set of bosonic baths, Eq.~(\ref{eq:H_int}) (see Fig.~\ref{fig:fig2}).
In order to study the dynamics we derived a kinetic equation, within
the Keldysh formalism. A detailed analysis of the characteristic relaxation
time obtained from the kinetic equation was given in Sec.~\ref{sec:Relaxation-time}.
An analytic expression for the critical relaxation time was obtained
in Eq.~(\ref{eq: tau diagonal}) and verified in Fig.~\ref{fig: tau s-gamma}.
As shown in Appendix B, the scaling of the latter as a function of
the temperature is related to the critical exponents of the model
(see Eq.~(\ref{eq:tau general appendix})).
Finally, we considered the quench dynamics. The kinetic equations derived allow us to
study the dissipative dynamics also beyond the universal regime. We checked the
scaling-laws derived and their range of validity in Figs.~\ref{fig:E and Eta} and \ref{fig:E hf}.

We remark that the method described here to obtain a kinetic
equation for the XY model, may be extended to describe the dissipative
dynamics of other models that can be mapped into fermionic degrees
of freedom, like other spin chains and ladders or certain $2d$ models
of the Kitaev-type.

\section*{Acknowledgments}

We acknowledge F. Guinea, V. Kravtsov, A. Polkovnikov, A.J. Leggett, R. Raimondi and F. Sols, T. Caneva,
G. Carleo and M. Schir\`o for fruitful discussions. D.P. acknowledges the ISTANS (grant 1758) program of
ESF for financial support.

\appendix

\section*{Appendix A1: kinetic equation} \label{appe1}

Here we present a detailed derivation of the kinetic equation.
Apart from the Keldysh (\ref{eq: Keldysh GF}), lesser (\ref{eq:lesser GF})
and greater (\ref{eq: greater GF}) Green's function we need also the
retarded and the advanced ones, and also the bath Green functions:
\begin{eqnarray*}
G_{kij}^{a(r)}(t_{1},t_{2}) &\doteq& (-)i \theta(t_{2(1)}-t_{1(2)})\left\langle
\left\{ \Psi_{ki}(t_{1}),\ \Psi_{kj}^{\dagger}(t_{2})\right\} \right\rangle \\
g_{q}(t_{1},t_{2}) &\doteq& -i \left\langle \mathcal{T}_{\gamma}\ X_{q}(t_{1})X_{q}(t_{2})\right\rangle \\
g_{q}^{<(>)}(t_{1},t_{2}) &\doteq& (-)i \left\langle \ X_{q}(t_{2(1)})X_{q}(t_{1(2)})\right\rangle \\
g_{q}^{a(r)}(t_{1},t_{2}) &\doteq& (-)i \theta(t_{2(1)}-t_{1(2)})
\left\langle \left[X_{q}(t_{1}),\ X_{q}(t_{2})\right]\right\rangle
\end{eqnarray*}
where we used commutators (anticommutators) for the retarded and advanced bath (system) Green's functions.
The starting point is the Dyson's equation (\ref{eq: Dyson eq}).
%
In order to obtain from the Dyson's equation (\ref{eq: Dyson eq}) an equation for the lesser
and greater Green's function we use the Keldysh book-keeping for a generic convolution
$C(t_{1},t_{2}) \doteq \int_{\gamma}d\bar{t} A(t_{1},\bar{t}) B(\bar{t},t_{2})$
is $C^{r(a)}(t_{1},t_{2}) = \int_{0}^{t}d\bar{t}A^{r(a)}(t_{1},\bar{t})B^{r(a)}(\bar{t},t_{2})$
and $C^{<(>)}(t_{1},t_{2}) = \int_{0}^{t}dt_{1} A^{r}(t_{1},\bar{t}) B^{<(>)}(\bar{t},t_{2})
+ A^{<(>)}(t_{1},\bar{t}) B^{a}(\bar{t},t_{2})$ \cite{vanLeeuwen05}.
Using the previous formulas we rewrite the Dyson's equations as:
\begin{eqnarray}
i \partial_{t_{1}}G_{k}^{<}(t_{1},t_{2}) & = & \mathcal{H}_{k}(t_{1})G_{k}(t_{1},t_{2})+\label{eq: G<}\\
 &  & \hspace{-1cm}\int_{0}^{t}d\bar{t}\ \Sigma_{k}^{r}(t_{1},\bar{t})G_{k}^{<}(\bar{t},t_{2})
          +\Sigma_{k}^{<}(t_{1},\bar{t})G_{k}^{a}(\bar{t},t_{2})\nonumber
\end{eqnarray}
and an analogous equation for $\partial_{t_{2}}$.
We are interested in the \emph{equal-time} Green's function and hence we perform a change
of variables:
\begin{eqnarray*}
t & = & \frac{t_{1}+t_{2}}{2}\\
\delta t & = & t_{1}-t_{2}
\end{eqnarray*}
whose Jacobian is simply $\partial_{t}=\partial_{t_{1}}+\partial_{t_{2}}$
and $\partial_{\delta t}=\frac{1}{2}(\partial_{t_{1}}-\partial_{t_{2}})$.
At equal time ($\delta t=0$), for the lesser Green's function $G_{k}^{<}=G_{k}^{<}(t_{1},t_{1})=G_{k}^{<}(t)$
we get:
\begin{eqnarray}
i \partial_{t}G_{k}^{<} & = & [\mathcal{H}_{k}(t),\ G_{k}]+\label{eq: G<2}\\
 &  & \Sigma_{k}^{r}\cdot G_{k}^{<}+\Sigma_{k}^{<}\cdot G_{k}^{a}
    -G_{k}^{r}\cdot\Sigma_{k}^{<}-G_{k}^{<}\cdot\Sigma_{k}^{a}\nonumber
\end{eqnarray}
where the dot indicates the convolution $\Sigma_{k}^{r}\cdot G_{k}^{<} \doteq
\int_{0}^{t}d\bar{t}\,\Sigma_{k}^{r}(t,\bar{t})G_{k}^{<}(\bar{t},t)$.
Now we use the relations:
\begin{eqnarray*}
G^{r}(t_{1},t_{2}) & = & \theta(t_{1}-t_{2})\left(G^{>}(t_{1},t_{2})-G^{<}(t_{1},t_{2})\right)\\
G^{a}(t_{1},t_{2}) & = & \theta(t_{2}-t_{1})\left(G^{<}(t_{1},t_{2})-G^{>}(t_{1},t_{2})\right)
\end{eqnarray*}
and similar relations that hold also for the $\Sigma^{r,a}$ (see \cite{vanLeeuwen05}):
\begin{eqnarray*}
\Sigma^{r(a)}(t_{1},t_{2}) & = & \Sigma^{\delta}\delta(t_{1},t_{2})+\\
 &  & \hspace{-1cm}\theta(t_{1(2)}-t_{2(1)})\left(\Sigma^{>(<)}(t_{1},t_{2})-\Sigma^{<(>)}(t_{1},t_{2})\right)
\end{eqnarray*}
where we can neglect the term $\Sigma^{\delta}$ that only renormalizes
the Hamiltonian and is not relevant in our case (see Eq.~(\ref{eq:SIGMA_lessgrt}) below).
At equal times we get:
\begin{eqnarray*}
i \partial_{t}G_{k}^{<} & = & \left[\mathcal{H}_{k},G_{k}^{<}\right]+\\
 &  & \Sigma_{k}^{>}\cdot G_{k}^{<}-\Sigma_{k}^{<}\cdot G_{k}^{>}
        +G_{k}^{<}\cdot\Sigma_{k}^{>}-G_{k}^{>}\cdot\Sigma_{k}^{<}
\end{eqnarray*}

We now perform a Markov approximation.
This will transform the integro-differential kinetic equation into a differential equation.
Let us define the interaction picture for a general function:
\begin{eqnarray*}
\tilde{O}_{k}(t_{1},t_{2}) &\doteq&
         \mathcal{U}_{k}^{\dagger}(t_{1})O_{k}(t_{1},t_{2})\mathcal{U}_{k}(t_{2})
\end{eqnarray*}
where $\mathcal{U}_{k}$ is the \emph{free} evolution matrix for the system obeying
$i \, \mathcal{\dot{U}}_{k} = \mathcal{H}_{k} \mathcal{U}_{k}$.
Such transformation gauges away the free evolution and the new Green's function $\tilde{G}_{k}$
dynamics is solely governed by the self energy:
\[
i \partial_{t}\tilde{G}_{k}^{<}=\tilde{\Sigma}_{k}^{>}\cdot\tilde{G}_{k}^{<}
      -\tilde{\Sigma}_{k}^{<}\cdot\tilde{G}_{k}^{>}
      +\tilde{G}_{k}^{<}\cdot\tilde{\Sigma}_{k}^{>}-\tilde{G}_{k}^{>}\cdot\tilde{\Sigma}_{k}^{<} \;.
\]
Since the self-energy carries the ``small'' perturbative coupling parameter
the evolution of $\tilde{G}_{k}$ can be regarded as ``slow'' with respect
to the time scale of the self energy, that is the same as that of the bath.
In fact $\Sigma$ contains the bath-correlation function $g(t_{1},t_{2})$
(see Eq.~(\ref{eq:SIGMA_lessgrt})) that is strongly peaked at $t_{1}\simeq t_{2}$
because of the assumption of a cutoff-time for the bosonic modes (see Sec.~\ref{sec:Kinetic-equation}).
Thus we can take $\tilde{G}$ out of the convolutions:
\begin{eqnarray}
i \partial_{t}\tilde{G}_{k}^{<} & \eqsim & \left(\int_{0}^{t} d\bar{t}
    \tilde{\Sigma}_{k}^{>}(t,\bar{t})\right) \tilde{G}_{k}^{<}
    -\left(\int_{0}^{t}d\bar{t}\tilde{\Sigma}_{k}^{<}(t,\bar{t})\right)\tilde{G}_{k}^{>}\nonumber \\
 &  & \hspace{-1.1cm}+\tilde{G}_{k}^{<}\left(\int_{0}^{t} d\bar{t}
        \tilde{\Sigma}_{k}^{>}(\bar{t},t)\right)-\tilde{G}_{k}^{>}
       \left(\int_{0}^{t}d\bar{t}\tilde{\Sigma}_{k}^{<}(\bar{t},t)\right)
\label{eq:KE_Markov_general}
\end{eqnarray}
Eq.~(\ref{eq:KE_Markov_general}) is quite general and it is based solely on the assumption of Markovian baths.
We now use the explicit form of the self-energy for the coupling system-bath (\ref{eq: H_int_NAMBU})
with $l=1$, that within the self-consistent Born approximation reads:
\begin{eqnarray}
\Sigma_{k}(t_{1},t_{2}) & = & \frac{i }{N}\sum_{q}\: g_{k-q}(t_{1},t_{2})\tau^{z}G_{q}(t_{1},t_{2})\tau^{z}\nonumber \\
 & = & \frac{i }{N}g(t_{1},t_{2})\tau^{z}\sum_{q}G_{q}(t_{1},t_{2})\tau^{z}
\label{eq:SIGMA_keldysh}
\end{eqnarray}
where $g_{q}(t_{1},t_{2})=-i \left\langle \mathcal{T}\ X_{q}(t_{1})X_{q}(t_{2})\right\rangle $
is the non-interacting bath Keldysh Green's function that does not explicitly
depend on the moment $q$ (since all baths have the same spectral function).
In Eq.~(\ref{eq:SIGMA_keldysh}) we neglected the polaronic shift contribution
(corresponding to the tadpole diagram, Fig.~\ref{fig: diagram}b)
\begin{equation}
\Sigma_{k}^{\delta}(t_{1},t_{2}) = -\frac{i }{N}\delta(t_{1,}t_{2})\tau^{z} \int_{\gamma}
d\bar{t}\, g(t_{1},\bar{t})\sum_{q}\textrm{Tr}[\tau^{z}G_{q}(\bar{t},\bar{t})]
\label{eq:SIGMA polaronic}
\end{equation}
In fact, being such term proportional to a $\delta(t_{1,}t_{2})$, it has only the irrelevant effect
of renormalizing the Hamiltonian (see Sec.~\ref{sec:Kinetic-equation}).
Using again the Keldysh book-keeping \cite{HaugBOOK,vanLeeuwen05},
we obtain from Eq.~(\ref{eq:SIGMA_keldysh}), for the lesser and greater self energy
\begin{eqnarray}
\Sigma_{k}^{\lessgtr}(t_{1},t_{2}) &=& \frac{i }{N }g^{\lessgtr}(t_{1},t_{2}) \tau^{z}
\sum_{q} G_{q}^{\lessgtr}(t_{1},t_{2})\tau^{z}
\label{eq:SIGMA_lessgrt}
\end{eqnarray}
(notice that $g^{>}(t_{1},t_{2})=-g^{<}(t_{1},t_{2})^{*}$).
Evaluating explicitly the self-energy kernels we obtain:
\begin{widetext}

\begin{eqnarray*}
\int_{0}^{t}d\bar{t} \tilde{\Sigma}_{k}^{>}(t,\bar{t}) &=&
\frac{i }{N}\sum_{q}\int_{0}^{t} d\bar{t}\ g^{>}(t,\bar{t}) \mathcal{U}_{k}^{\dagger}(t) \tau^{z}
 G_{q}^{>}(t,\bar{t})\tau^{z}\mathcal{U}_{k}(\bar{t})\\
 &=& \frac{i }{N}\sum_{q} \int_{0}^{t} d\bar{t\ } g^{>}(t-\bar{t}) \mathcal{U}_{k}^{\dagger}(t) \tau^{z}
    \mathcal{U}_{q}(t) \tilde{G}_{q}^{>}(t,\bar{t}) \mathcal{U}_{q}^{\dagger}(\bar{t}) \tau^{z}
    \mathcal{U}_{k}(\bar{t})\\
 &\simeq& \frac{i}{N} \sum_{q} \mathcal{U}_{k}^{\dagger}(t) \tau^{z} \mathcal{U}_{q}(t) \tilde{G}_{q}^{>}(t,t)
  \left(\int_{0}^{\infty}\! ds\ g^{>}(s)\mathcal{U}_{q}^{\dagger}(t-s) \tau^{z} \mathcal{U}_{k}(t-s)\right)
\end{eqnarray*}
\begin{eqnarray*}
\int_{0}^{t} d\bar{t} \tilde{\Sigma}_{k}^{>}(\bar{t},t) &=&
\frac{i }{N}\sum_{q}\int_{0}^{t}d\bar{t}\ g^{>}(\bar{t},t)\mathcal{U}_{k}^{\dagger}(\bar{t}) \tau^{z}
G_{q}^{>}(\bar{t},t) \tau^{z} \mathcal{U}_{k}(t)\\
 &=& \frac{i}{N} \sum_{q} \int_{0}^{t} d\bar{t}\ g^{>}(\bar{t}-t)\mathcal{U}_{k}^{\dagger}(\bar{t}) \tau^{z}
 \mathcal{U}_{q}(\bar{t})\tilde{G}_{q}^{>}(\bar{t},t) \mathcal{U}_{q}^{\dagger}(t) \tau^{z} \mathcal{U}_{k}(t)\\
 &\simeq& \frac{i}{N} \sum_{q} \left(\int_{0}^{\infty} \! ds\ g^{>}(-s)_{q} \mathcal{U}_{k}^{\dagger}(t-s)
    \tau^{z} \mathcal{U}_{q}(t-s)\right) \tilde{G}_{q}^{>}(t,t) \mathcal{U}_{q}^{\dagger}(t)
    \tau^{z} \mathcal{U}_{k}(t)
\end{eqnarray*}
\end{widetext}
where $\simeq$ refers again to the Markov approximation.
For the greater kernels simply interchange ``$<$'' with `$>$''.
Finally, in the Schr\"odinger picture, using the relation $G^{>}=-i{\bf 1}+G^{<}$, we
obtain Eq.~(\ref{eq:Kinetic}).

\section*{Appendix A2: Approximation for the kinetic equation matrices $\hat{D}$} \label{appe2}

In this appendix we comment on the validity of the approximation (\ref{eq: Dapprox main})
for the matrices (\ref{eq: D matrices main}) appearing in the kinetic equation.
To calculate $\hat{D}$ exactly, we need to know the evolution operator $\mathcal{\hat{U}}_{k}$,
solution of the differential equation $i\, \mathcal{\dot{\hat{U}}}_{k}=\mathcal{\hat{H}}_{k}\mathcal{\hat{U}}_{k}$.
This can be obtained exactly by mapping the dynamics of a generic mode $k$ into a Landau-Zener two-level
system dynamics \cite{Dziarmaga05,Fubini07}
\begin{equation}
\mathcal{\hat{H}}^{LZ} \equiv h^{LZ}(t) \hat{\tau}^{z}+\Delta^{LZ} \hat{\tau}^{x}
\label{eq:H_LZ}
\end{equation}
with $\Delta_{k}^{LZ}=\gamma\sin k$ and $h_{k}^{LZ}=-vt$ that can be obtained from (\ref{eq: H matrix Nambu})
by a simple change of the variable $t$. Using the solution of the Landau-Zener problem
for a quench that starts at $t=-\infty$ we have for the matrix elements of $\mathcal{U}^{LZ}(-\infty,t)$,
\[
\mathcal{U}^{LZ}(-\infty,t)=\left(\begin{array}{cc}
\mathcal{U}_{11}^{LZ}(t) & -\mathcal{U}_{21}^{LZ}(t)^{*}\\
\mathcal{U}_{21}^{LZ}(t) & \mathcal{U}_{11}^{LZ}(t)^{*}\end{array}\right)
\]
the following results:
\begin{eqnarray*}
\mathcal{U}_{11}^{LZ}(t) &=& e^{i \frac{\pi}{4}}\exp\left\{ -\frac{\pi\left(\Delta^{LZ}\right)^{2}}{8v}\right\}
      \mathcal{D}_{-p}\left((-1+i )\sqrt{v}\ t\right)\\
\mathcal{U}_{21}^{LZ}(t) &=& \frac{\Delta^{LZ}}{\sqrt{2v}}\exp\left\{ -\frac{\pi\left(\Delta^{LZ}\right)^{2}}{8v}
      \right\} \mathcal{D}_{-p-1}\left((-1+i )\sqrt{v}\ t\right)
\end{eqnarray*}
where $p=-i \left(\Delta^{LZ}\right)^{2}/2v$ and $\mathcal{D}_{p}$
are parabolic cylinder functions. Finally the evolution operator from
generic $\bar{t}$ to $t$ can be obtained using the simple property:
\begin{eqnarray*}
\mathcal{U}^{LZ}(\bar{t},t) &=& \mathcal{U}^{LZ}(-\infty,t)\mathcal{U}^{LZ}(\bar{t},-\infty)\\
 &=& \mathcal{U}^{LZ}(-\infty,t)\mathcal{U}^{LZ\dagger}(-\infty,\bar{t})
\end{eqnarray*}

The second ingredient we need in order to calculate $\hat{D}$ is
the bath thermal equilibrium correlation function $g^{>}$.
From its definition:
\begin{eqnarray*}
g^{>}(t) &\doteq& -i \left\langle X(t)X^{\dagger}(0)\right\rangle \\
 & = & -i \sum_{\beta}\lambda_{\beta}^{2}
     \left( e^{-i\omega_{\beta}t} \left\langle b_{-\beta}b_{-\beta}^{\dagger} \right\rangle
          + e^{ i\omega_{\beta}t} \left\langle b_{\beta}^{\dagger}b_{\beta}\right\rangle \right)\\
 & = & -i \int_{0}^{\infty}\! d\omega \, J(\omega) [e^{-i\omega t}(1+n_{B}(\omega))+e^{i\omega t} n_{B}(\omega)]
\end{eqnarray*}
where $n_{B}\equiv 1/(e^{\omega/T}-1)$ is the Bose function and we used the definition
(\ref{eq:Bath Spectral Function}) of spectral function for the bath
$J(\omega)=\sum_{\beta}\lambda_{\beta}^{2}\delta(\omega-\omega_{\beta})$.
The correlation function can be written explicitly as:
\begin{eqnarray*}
g^{>}(t) &=& -i \int_{0}^{\infty}\! d\omega \, J(\omega)
\left(\coth(\frac{\omega}{2T})\cos(\omega\tau)-i\sin(\omega\tau) \right)\\
 & = & -2i\alpha T^{s+1} \Gamma(s+1) \times\\
 &  & \hspace{-1cm} \left( \zeta(s+1,\, T\frac{1+\frac{\omega_{c}}{T}-i\omega_{c}\tau}{\omega_{c}})
                          +\zeta(s+1,\, T\frac{1+i\omega_{c}\tau}{\omega_{c}})\right)
\end{eqnarray*}
where $\Gamma$ is the Gamma function and $\zeta(z,u)\equiv\sum_{n=0}^{\infty}\frac{1}{(n+u)^{z}},\; u\neq 0,-1,-2\dots$.

\begin{figure}[h]
\includegraphics[scale=0.7]{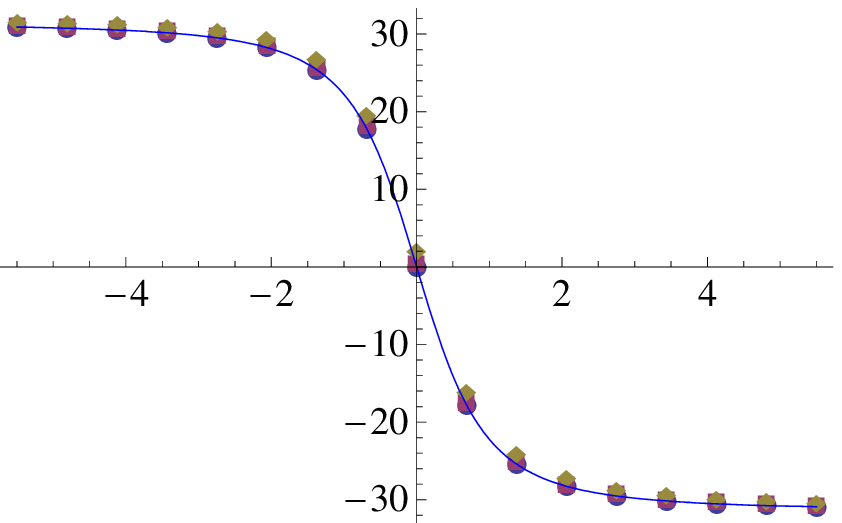}\\
 \includegraphics[scale=0.7]{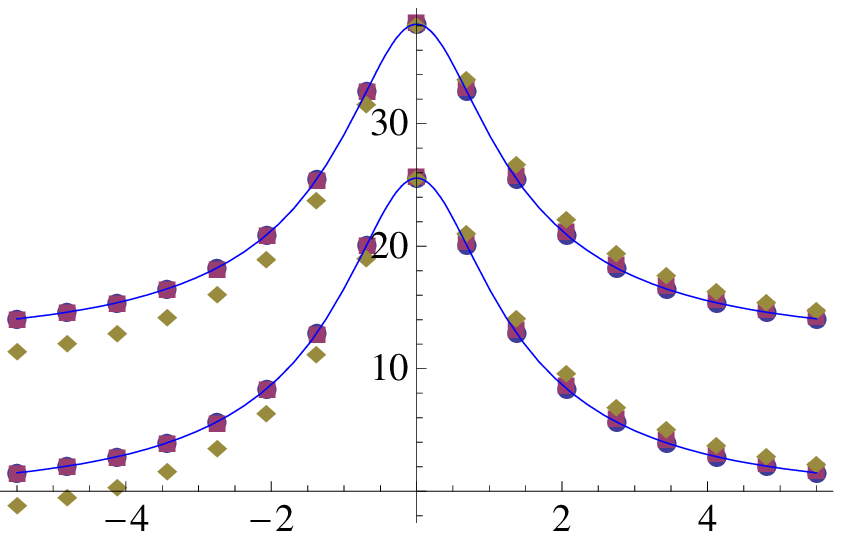}
\label{fig:Delements}
\caption{Matrix elements of $\hat{D}_{kk}$ as a function of the rescaled field
$h^{LZ}$ (\ref{eq:H_LZ}); $h^{LZ}=0$ correspond to the critical
point $h\simeq1$ for the relevant low energy modes. Lower panel:
diagonal elements $i\left(D_{kk}\right)_{21}$ (up) and $-i\left(D_{kk}\right)_{12}$
(down); upper panel shows the difference $\left(D_{kk}\right)_{11}-\left(D_{kk}\right)_{22}$.
Plots refer to $T/\Delta^{LZ}=5$; continuous line is the approximation
(\ref{eq: Dqk}) (which is independent on $v$) and symbols are the
exact value for $\sqrt{v}/T=0.1,\ 1,\ 5$; deviations from the approximation
are appreciable only for the last value of $v$. }
\end{figure}

We are now able to calculate explicitly the matrix $\hat{D}$ and
check the validity of the approximation in (\ref{eq: Dapprox main}).
As stated in Sec.~\ref{sec:Kinetic-equation}, the approximation
consists in considering the instantaneous transition rates induced
by the bath, independent on the velocity of the quench.
This is ultimately justified by the assumption of ``fast'' and memoryless Markovian baths.
Hence, for slow quenches when the typical time of the quench
is much larger than the typical bath time-scale we expect that the
magnetic field can be regarded as not evolving for the bath.
Within the approximation in Eq.~(\ref{eq: Dapprox main}) we can perform explicitly
the integration for the matrix elements of $\hat{D}_{qk}$ over time,
giving the Laplace transform of the bath Green function.
We are interested only in the real part of the latter
\begin{eqnarray}
g[E] &\doteq& \Re[ \int_{0}^{\infty} i g^{>}(t) e^{i Et}] \label{eq:g[E]} \\
     &=& \pi \left( J(E)-J(-E)\right) \frac{\exp\left( \beta E\right) }{\exp\left( \beta E\right)-1}
\nonumber
\end{eqnarray}
since the imaginary part gives a renormalization contribution that
is negligible in the weak coupling limit $\alpha\rightarrow 0$ \cite{CohenBOOK}.
In the basis of the eigenvectors of $\mathcal{\hat{H}}_{k}$ we obtain:
\begin{widetext}
\begin{equation}
\hat{D}_{qk}^{approx} = \frac{1}{2} \left(\begin{array}{cc}
\cos^{++}g^{+-}+\cos^{+-}g^{++} & i\left(\sin^{+-}g^{-+}+\sin^{++}g^{--}\right)\\
-i\left(\sin^{+-}g^{+-}+\sin^{++}g^{++}\right) & \cos^{-+}g^{--}-\cos^{++}g^{-+}\end{array}\right) \;.
\label{eq: Dqk}
\end{equation}
\end{widetext}
where we defined
\begin{eqnarray*}
\cos^{\pm\pm} &\doteq& \pm\cos\theta_{k}\pm\cos\theta_{q}\\
\sin^{\pm\pm} &\doteq& \pm\sin\theta_{k}\pm\sin\theta_{q}\\
g^{\pm\pm}    &\doteq& g[\pm\Lambda_{k}\pm\Lambda_{q}]
\end{eqnarray*}
The $\sin$ and $\cos$ are geometric factors specific of the system
operator that couples to the bath (in our case $\sigma^{z}$), while
the Laplace transform of the bath $g[.]$ (see \ref{eq:g[E]}) carries information about
the relaxation rates between the different energy levels, and depends
explicitly on the temperature and on the nature of the baths
(i.e., its spectral function).
For simplicity we consider the equal indexes $\hat{D}_{kk}$ matrices.
The same results hold also for unequal indexes since the integral of the matrix elements
have the same structure in both cases. In Fig.~\ref{fig:Delements}
we compare the matrix elements obtained using the exact evolution
operator with the ones given by Eq.~(\ref{eq: Dqk}).
The agreement is good, validating the approximation. Deviations appear only in the
limit of fast quenches $\sqrt{v}\gg T$, and in such regime the bath
has a less relevant effect on the dynamics because of the short interaction
time during the quench. Besides that, deviations appear far from the critical
point (corresponding to $h^{LZ}\simeq 0$), i.e., far from the most
relevant part of the quench according to Secs.~\ref{sec:Scaling-analysis}
and \ref{sec:Adiabatic-quenches}.

\section*{Appendix B: Fermi golden rule for the relaxation time} \label{appe3}

In this section we derive an expression for the critical relaxation
time using the Fermi golden rule for a generic system interacting
with a bosonic bath. Let us assume the system-bath interaction Hamiltonian to have the form
$H_{int}=AZ$ where $A$ and $Z$ are system and bath operator respectively.
%
Consider a quench of the system from zero temperature to a certain finite $T$.
The transition rate for the process of thermalization in presence
of the reservoir $\rho_{B}^{th} \otimes (|GS\rangle\langle GS|)_{S}\rightarrow \rho_{B}^{th} \otimes \rho_{S}^{th}$
(where $B$ and $S$ refer to system and bath, respectively) is:
\begin{eqnarray}
\frac{1}{\tau} &=& 2\pi\sum_{f,i,k} \delta(E_{f}+E_{k}-E_{i}-E_{GS})P_{B}^{th}(E_{i}/T) \nonumber \\
 & & \times P_{S}^{th}(E_{k}/T)|\left\langle k,\ f\right|H_{int}\left|GS,\ i\right\rangle |^{2} \label{eq:FGR 1}
\end{eqnarray}
where $i$, $f$ and $k$ address the bath eigenvalues and the final state of the system respectively;
$P_{S(B)}^{th}$ are thermal weights.
We rewrite the $\delta$-function as
$\frac{1}{2\pi}\int_{-\infty}^{\infty}dt e^{-i(E_{f}-E_{i})t} e^{-i(E_{k}-E_{GS})t}$.
Summing over $f$ and $i$ we get the bath correlation function
$z(t)=\left\langle Z(t)Z(0)\right\rangle$:
\begin{equation}
\frac{1}{\tau} = \sum_{k}\int_{-\infty}^{\infty}dt e^{-i(E_{k}-E_{GS})t} z(t) P_{S}^{th}(E_{k}/T)|
     \left\langle k\right|A\left|GS\right\rangle |^{2} \label{eq:FGR 2}
\end{equation}
The time integral gives the Fourier transform of the bath correlation function, that we parametrize as
\begin{equation}
z[E] = J(E) f(E/T)  \;.
\label{eq:FT g}
\end{equation}
For instance, for a bosonic bath with spectral function $J(E)\propto E^{s}$ we have
\[
z[E]=\begin{cases}
J(E)\, (1+n_{B}(E/T)) & E>0\\
J(|E|)\, n_{B}(|E|/T)     & E<0
\end{cases} \;.
\]
By integrating over the k-modes (setting $E_{GS}=0$), using the critical density of states
$\rho(E)\propto E^{d/z-1}$, we get:
\begin{eqnarray}
\frac{1}{\tau}
 &\propto& \int \! dE \, \rho(E) \, z[E] \, P_{S}^{th}(E/T)\, |A_{GS}(E)|^{2} \;, \nonumber
\end{eqnarray}
where $A_{GS}(E)=\left\langle k(E)\right|A\left|GS\right\rangle$.
Now, assuming that the low-energy modes (that are the relevant ones at low-temperature) are coupled uniformly
by the bath $A_{GS}(E)\simeq A_{GS}(0)$ we obtain
\begin{equation}
\frac{1}{\tau} \propto \int \! dE\, E^{d/z-1}\ J(E) \, f(E/T) \, P_{S}^{th}(E/T)
\label{eq:tau E0}
\end{equation}
and finally, using for the spectral density $J(E)\propto E^{s}$
and performing a change of variable to $x=E/T$ we obtain
\begin{equation}
\tau^{-1}\propto T^{s+d/z} \;.
\label{eq:tau general appendix}
\end{equation}



\begin{thebibliography}{10}
\bibitem{footnote1}
    Notice that depending on the characteristics of the bath, the system-bath coupling may lead to a
    change of universality class of the transition as a result of bath
    induced long-range correlations effects. We will not address these
    issues here, which deserve a careful separate study.
    Therefore we will assume the bath correlations to be short ranged in time \protect\cite{PatanePRL}.
\end{thebibliography}
\end{document}